\pdfoutput=1
\documentclass[PAPER, coverpage=false, cernpreprint, atlasdraft=false, UKenglish, texmf, orcidlogo]{atlasdoc}
\usepackage{atlaspackage}
\usepackage{atlasbiblatex}

\usepackage{atlasphysics}
\usepackage{tabularht}

\usepackage{tikz}
\usetikzlibrary{decorations.pathmorphing}
\usetikzlibrary{decorations.markings}

\graphicspath{{logos/}{figures/}}

\usepackage{ANA-EXOT-2022-42-PAPER-defs}

\AtlasTitle{A measurement of the high-mass $\tau\bar{\tau}$ production cross-section at $\sqrt{s}=13$~TeV with the ATLAS detector
and constraints on new particles and couplings}

\AtlasAbstract{%
The production cross-section of high-mass $\tau$-lepton pairs is measured as a function of the dilepton visible invariant mass,
using 140 fb$^{-1}$ of $\sqrt{s}=13$~TeV proton--proton collision data recorded with the ATLAS detector at the Large Hadron
Collider.  The measurement agrees with the
predictions of the Standard Model.  A fit to the invariant mass distribution is performed as a function of $b$-jet multiplicity, to
constrain the non-resonant production of new particles described by an effective field theory or in models containing leptoquarks or
$Z'$ bosons that couple preferentially to third-generation fermions.  The constraints on new particles improve on previous results,
and the constraints on effective operators include those affecting the anomalous magnetic moment of the $\tau$-lepton.
}

\PreprintIdNumber{CERN-EP-2025-063}

\AtlasJournalRef{\JHEP 10 (2025) 54}
\AtlasDOI{10.1007/JHEP10(2025)054}

\hypersetup{pdftitle={ATLAS document},pdfauthor={The ATLAS Collaboration}}

\begin{document}

\maketitle

\tableofcontents

\section{Introduction}
\label{sec:intro}
The study of Drell--Yan production of charged-lepton pairs resulted in the historic discoveries of the
$J/\psi$~\cite{E598:1974sol,SLAC-SP-017:1974ind} and $\Upsilon$~\cite{E288:1977xhf} mesons, and of the
$Z$ boson~\cite{UA1:1983mne,UA2:1983mlz}.  Further studies have provided precise measurements of the
weak mixing angle \cite{CMS-SMP-22-010,STDM-2011-34,CDF:2018cnj} and constraints on parton distribution
functions~\cite{STDM-2019-18,CMS-SMP-13-003,D0:2007djv,CDF:2010vek}.
At high invariant mass, Drell--Yan production probes a variety of possible new interactions including the
production or exchange of $Z'$ bosons \cite{EXOT-2018-08,CMS-EXO-19-019}, leptoquarks~\cite{Haisch:2022afh},
and additional Higgs bosons~\cite{HDBS-2018-46}. There are well-motivated scenarios for these particles
to have enhanced couplings to third-generation fermions~\cite{Davighi:2023iks}, making Drell--Yan production
in the $\tau\bar{\tau}$ final state a promising probe for new physics.  For example, measurements of
$B$-hadron decays to $D^{(*)}\tau\nu$ show increased rates relative to the corresponding decays to lighter
leptons~\cite{Capdevila:2023yhq,HFLAV:2022esi,LHCb:2015gmp,BaBar:2013mob,Belle:2015qfa,Belle:2017ilt,
Belle:2019rba,LHCb:2017rln,LHCb:2023uiv}.  This could be explained by the presence of a leptoquark or
$Z'$ boson that preferentially couples to third-generation fermions and would affect $\tau\bar{\tau}$
production at high invariant mass.

Prior studies of $pp \to \tau\bar{\tau}$ production have probed the $Z$ boson resonance
\cite{LHCb:2018ogb,CMS-EWK-10-013,STDM-2011-18}, while studies of light leptons have extended to
high invariant mass~\cite{STDM-2014-06,STDM-2012-10,CMS-SMP-17-001,CMS-SMP-14-003,CMS-SMP-13-003,CMS-EWK-10-007}.
Searches for new particles have probed the high-mass region for both resonant and non-resonant production
of $\tau$-lepton pairs \cite{CMS-SMP-14-003,HIGG-2016-12,CMS-EXO-16-008,EXOT-2022-39,CMS-EXO-19-016}.  This paper
presents the first measurement of the inclusive $\tau\bar{\tau}$ fiducial cross-section as a function of visible
invariant mass, along with a search for new states through non-resonant interactions.  These interactions include
generic four-fermion interactions and $\tau\tau\gamma$ interactions that affect the anomalous magnetic moment of
the $\tau$-lepton.  The analysis is performed using the hadronic decays of the $\tau$-leptons since the
Drell--Yan purity is lower when there is one or more leptonic decays.

To improve sensitivity to third-generation couplings, non-resonant interactions are probed in bins
of $b$-jet multiplicity.  A $b$-quark from the proton participating in the hard interaction will frequently
have an associated $\bar{b}$ quark with substantial momentum.  The maximum sensitivity to new third-generation
interactions is observed in final states with one $b$-jet.  The fit in bins of $b$-jet multiplicity constrains
significant backgrounds, improving the sensitivity to new processes.  For interactions beyond the SM (BSM), the
BSM-only process is considered along with its interference with the SM production of $\tau$-lepton pairs.  The
interference contribution is found to significantly affect signal constraints, and has not been included in
previous non-resonant searches in this final state~\cite{EXOT-2022-39,CMS-EXO-19-016}.

This paper is organized as follows.  Section~\ref{sec:detector} describes the ATLAS detector and the data samples.
Section~\ref{sec:tauid} gives a brief overview of the reconstruction of physics objects, focusing primarily on
the reconstruction and identification of hadronic $\tau$-lepton decays.  A detailed discussion of the data
modelling follows in Section~\ref{sec:modelling}, covering both the Monte Carlo samples used in the analysis and
the data-driven estimate of the misreconstructed $\tau$-lepton background.  The results of the $\tau\bar{\tau}$
cross-section measurement are presented in Section~\ref{sec:xsec}, and those of the non-resonant search in
Section~\ref{sec:search}.  Conclusions are provided in Section~\ref{sec:conclusions}.

\section{ATLAS detector and data set}
\label{sec:detector}
The ATLAS experiment~\cite{PERF-2007-01} at the Large Hadron Collider (LHC)~\cite{Evans:2008zzb} is a
multipurpose particle detector with a forward--backward symmetric cylindrical geometry and a near \(4\pi\)
coverage in solid angle.\footnote{\AtlasCoordFootnote}
It consists of an inner tracking detector (ID) surrounded by a thin superconducting solenoid providing
a \qty{2}{\tesla} axial magnetic field, electromagnetic and hadronic calorimeters, and a muon spectrometer (MS).
The ID covers the pseudorapidity range \(|\eta| < 2.5\).  It consists of silicon pixel, silicon microstrip, and
transition radiation tracking detectors.  Lead/liquid-argon (LAr) sampling calorimeters provide electromagnetic
(EM) energy measurements with high granularity within the region \(|\eta|< 3.2\).  A steel/scintillator-tile
hadronic calorimeter covers the central pseudorapidity range (\(|\eta| < 1.7\)).  The endcap and forward regions
are instrumented with LAr calorimeters for EM and hadronic energy measurements up to \(|\eta| = 4.9\).
The MS surrounds the calorimeters and is based on three large superconducting air-core toroidal
magnets with eight coils each.  The field integral of the toroids ranges between \num{2.0} and
\qty{6.0}{\tesla\metre} across most of the detector.  The MS includes a system of precision tracking chambers up
to \(|\eta| = 2.7\) and fast detectors for triggering up to \(|\eta| = 2.4\).

A two-level trigger system is used to select events~\cite{TRIG-2016-01}.  The first-level trigger is implemented
in hardware and uses a subset of the detector information to accept events at a rate below \qty{100}{\kHz}.  This
is followed by a software-based trigger that reduces the accepted event rate to \qty{1}{\kHz} on average depending
on the data-taking conditions.  A software suite~\cite{SOFT-2022-02} is used in data simulation, in the
reconstruction and analysis of real and simulated data, in detector operations, and in the trigger and data
acquisition systems of the experiment.

The data set was collected between 2015 and 2018, during the LHC Run-2 production of $\sqrt{s}=13$~TeV $pp$ collisions,
and corresponds to an integrated luminosity of 140~fb$^{-1}$ with an uncertainty of 0.83\% \cite{DAPR-2021-01}.  The
LUCID-2 detector~\cite{LUCID2} is used for the primary luminosity measurements, which are complemented by measurements
using the inner detector and calorimeters.  Events are included from runs with fully operational tracking detectors and
calorimetery~\cite{DAPR-2018-01}, and were primarily collected with a trigger requiring a hadronically decaying $\tau$-lepton
with $\pt > 80$~GeV and another with $\pt > 60$~GeV in the software-based trigger. A single-$\tau$ trigger contributes
significantly to the efficiency at high invariant mass.  To ensure a high trigger efficiency the
offline selection requires the highest \pt $\tau$-lepton to have $\pt > 90$~GeV.  The selection yields a total of 6679
events.  Additional validation data sets were primarily collected with a trigger requiring a single electron or muon with
$\pt > 28$~GeV.  Events from these triggers, along with those from a single-jet trigger, are used in the background
estimation~\cite{ATLAS:2025akg}.

\section{Object reconstruction}
\label{sec:tauid}
This analysis uses events with two reconstructed $\tau$-leptons decaying into hadrons ($\tau_{\textrm{had}}$).
Validation of the modelling is performed using events with one $\tau_{\textrm{had}}$ and one charged lepton ($e$ or $\mu$).
The data are separated into bins of $b$-jet multiplicity to maximize sensitivity to interactions involving $b$-quarks.
Standard ATLAS procedures are applied for the reconstruction and calibration of these objects.

Hadronic $\tau$-lepton decays produce a neutrino and visible decay products, typically one or three charged pions
and up to two neutral pions. The reconstruction of the $\tau_{\textrm{had}}$ object is seeded by jets reconstructed
using the anti-$k_t$ algorithm~\cite{Cacciari:2008gp} with a radius parameter of $R = 0.4$. Reconstructed tracks are
then matched to $\tau_{\textrm{had}}$ candidates.
These candidates must have one or three associated tracks whose charge sum is $\pm 1$, and must have
$|\eta| < 2.47$ excluding the transition region between the barrel and endcap calorimeters ($1.37 < |\eta| < 1.52$)
due to the large amount of material in front of the first active layer.  In order to separate the $\tau_{\textrm{had}}$
candidates produced by hadronic $\tau$-lepton decays from those produced by jets initiated by quarks or gluons,
a recurrent neural-network identification algorithm (\RNN)~\cite{ATL-PHYS-PUB-2019-033} is used. This algorithm uses
information from reconstructed charged-particle tracks and calorimeter energy clusters,
together with high-level discriminating variables. The $\tau_{\textrm{had}}$ objects are required to fulfill the `Tight'
identification criterion. A separate discriminant (‘eBDT’)~\cite{ATL-PHYS-PUB-2015-045} is constructed to reject
backgrounds originating from electrons misidentified as $\tau_{\textrm{had}}$, arising mostly from $Z \rightarrow ee$
events in this analysis. In events with an identified electron, the $\tau_{\textrm{had}}$ object is required to pass
the `Loose' working point of the eBDT algorithm.  Systematic uncertainties in the $\tau_{\textrm{had}}$ reconstruction
vary with \pt and range from 2-6\% for the RNN and 1-2\% for the eBDT.  The $\tau_{\textrm{had}}$ energy scale is
studied using $Z\to \tau\tau$ decays in events with one muon and one $\tau_{\textrm{had}}$, and has an uncertainty of
1-4\%~\cite{ATLAS-CONF-2017-029}.

Jets are reconstructed using the anti-$k_{\textrm{T}}$ algorithm as implemented in
FastJet~\cite{Fastjet} with a jet radius parameter R = 0.4. The inputs to this algorithm are particle flow (PFlow)
objects~\cite{PERF-2015-09}, which combine measurements from the ATLAS inner detector and calorimeters~\cite{PERF-2014-07}
to improve the jet energy resolution and increase the jet reconstruction efficiency, especially at low jet \pt.
A jet vertex tagger (JVT)~\cite{PERF-2014-03} is used to remove jets with $\pt < 60$~GeV and $|\eta| < 2.5$ that are
identified as not being associated with the event’s primary vertex, defined as the vertex with the highest $\sum \pt^2$
of associated tracks.  Similarly, pile-up jets in the forward region are
suppressed with a forward JVT~\cite{PERF-2016-06} algorithm, which exploits jet shapes and topological jet correlations
in pile-up interactions and is applied to all jets with $\pt < 60$ GeV and $|\eta| > 2.5$.
Calibrations are applied to the jet energy scale (JES) and jet energy resolution (JER), which include components
derived both from simulation and in situ measurements~\cite{PERF-2016-04}. Systematic uncertainties in the
JES and JER are evaluated using a series of simulation-based techniques and in-situ measurements.

Jets originating from $b$-quarks are identified using the DL1r $b$-tagging algorithm~\cite{FTAG-2019-07} and are required
to have $\pt > 25$~GeV and $|\eta| < 2.5$.  Operating points are defined by a single threshold on the discriminant
output distribution and are chosen to provide a specific \bjet efficiency on an inclusive $t\bar{t}$ Monte Carlo
sample~\cite{FTAG-2018-01}.  The 77\% working point is used to select $b$-jets.
Correction factors are applied to the simulated
event samples to compensate for differences between data and simulation in the $b$-tagging efficiency for $b$, $c$ and
light-quark jets. The correction for \bjets is derived from $t\bar{t}$ events with final states containing two leptons,
and the corrections are consistent with unity with uncertainties at the level of a few percent over most of the jet \pt
range.

Electrons are reconstructed from topological clusters of energy deposits in the electromagnetic calorimeter, matched to
a track reconstructed in the ID. They are required to satisfy the `Medium' identification criteria~\cite{EGAM-2021-01},
to have $\pt > 7$~GeV, and to be in the fiducial volume of the ID and the high-granularity electromagnetic
calorimeters, $|\eta| < 2.47$, with the exclusion of the LAr barrel–endcap transition region $1.37 < |\eta| < 1.52$.
The Medium identification has an efficiency of 80\% to 90\%.  The clusters are additionally required to pass the
`FCHighPTCaloOnly' isolation criteria~\cite{EGAM-2018-01}.

Muons are reconstructed from signals in the MS matched with tracks in the ID. They are required to satisfy the `Medium'
identification criteria and the `FCLoose' isolation criteria~\cite{PERF-2015-10}, and to have $\pt > 7$~GeV
and $|\eta| < 2.5$.  The Medium identification has an efficiency above 97\%.

Since the objects used in the analysis are reconstructed from the same set of tracks and calorimetric energy clusters,
some constituents can be associated with multiple objects. These overlaps between different objects are resolved with
the procedure described in Ref.~\cite{HIGG-2019-09}.

\section{Modelling}
\label{sec:modelling}
The data are modelled using Monte Carlo (MC) event samples corresponding to the dominant Drell--Yan process and the
top-quark processes that contribute significantly to events containing one or more $b$-jets.  The small diboson and
Higgs boson components are also modelled using MC.  A final component consists of one or two hadronic jets
misreconstructed as $\tau_{\textrm{had}}$ objects, and is evaluated using data-driven techniques.

\subsection{Monte Carlo samples}

The Drell--Yan production of %
lepton pairs is modelled using \SHERPA[2.2.11]~\cite{Bothmann:2019yzt} events
generated with matrix-element calculations at next-to-leading order (NLO) in QCD for up to two additional partons,
and calculations at leading order (LO) for up to five additional partons using the {\textsc{Comix}}~\cite{Gleisberg:2008fv}
and \OPENLOOPS~\cite{Buccioni:2019sur,Cascioli:2011va,Denner:2016kdg} libraries.  The calculations are matched to the
\SHERPA parton shower based on Catani–Seymour dipole factorisation~\cite{Gleisberg:2008fv,Schumann:2007mg} using
the \MEPSatNLO prescription~\cite{Hoeche:2011fd,Hoeche:2012yf,Catani:2001cc,Hoeche:2009rj}.  The \NNPDF[3.0nnlo]
set of parton distribution functions (PDFs)~\cite{Ball:2014uwa} are used.  Uncertainties in the Drell--Yan
predictions are estimated by comparing distributions to those from events generated with the
\POWHEGBOX[v1]~\cite{Nason:2004rx,Frixione:2007vw,Alioli:2010xd,Alioli:2008gx} generator at NLO accuracy in QCD and
interfaced to \PYTHIA[8.186]~\cite{Sjostrand:2007gs} for the modelling of the parton shower, hadronisation, and underlying
event, with parameters set according to the AZNLO set of tuned parameters (tune)~\cite{STDM-2012-23}.  For these events,
the \CT[10nlo] PDF set~\cite{Lai:2010vv} is used for the hard-scattering process, and the \CTEQ[6L1] PDF
set~\cite{Pumplin:2002vw} is used for the parton shower.  The effect of QED final-state radiation was simulated with
\PHOTOSpp[3.52]~\cite{Golonka:2005pn,Davidson:2010ew}, and the \EVTGEN[1.2.0] program~\cite{Lange:2001uf} is used to
decay bottom and charm hadrons.

The production of $t\bar{t}$ events is modelled using the \POWHEGBOX[v2] generator at NLO in QCD with the \NNPDF[3.0nlo]
PDF set and the $h_{\textrm{damp}}$ parameter\footnote{The $h_{\textrm{damp}}$ parameter is a resummation damping factor
and one of the parameters that controls the matching of \POWHEG matrix element to the parton shower and thus
effectively regulates the high-\pt radiation against which the $t\bar{t}$ system recoils.} set to
1.5\,\mtop~\cite{ATL-PHYS-PUB-2016-020}. The events are interfaced to \PYTHIA[8.230]~\cite{Sjostrand:2014zea} to model
the parton shower, hadronisation, and underlying event, with parameters set according to the A14
tune~\cite{ATL-PHYS-PUB-2014-021} and using the \NNPDF[2.3lo]~\cite{Ball:2012cx} set of PDFs. The decays of bottom and
charm hadrons are performed by \EVTGEN[1.6.0].  Parton-shower uncertainties are estimated by replacing the \PYTHIA parton
shower with that of \HERWIG[7.13]~\cite{Bahr:2008pv,Bellm:2015jjp} using the default set of tuned parameters and the
MMHT2014 PDF set~\cite{Harland-Lang:2014zoa}.  The uncertainty in the matching of the matrix element (ME) to the parton
shower is estimated using events generated with the \PYTHIA `POWHEG::pThard=1' option, which changes the scale from the
minimum \pt of all final-state particles in the \POWHEG calculation to the \pt of the \POWHEG emission.

Individual top quarks can be produced in association with a $W$ boson, in the $t$-channel, or in the $s$-channel.  Single-top
processes are modelled using the \POWHEGBOX[v2] generator at NLO in QCD using the \NNPDF[3.0nlo] PDF set, interfaced
to \PYTHIA[8.230] using the A14 tune and the \NNPDF[2.3lo] PDF set.  The four-flavour (five-flavour) scheme is used for
$t$-channel ($s$-channel and associated) production.  For associated production the diagram subtraction
scheme~\cite{Frixione:2008yi,ATL-PHYS-PUB-2016-020} is used to remove interference and overlap with $t\bar{t}$ production.
The related uncertainty is estimated by comparing with an alternative sample generated using the diagram removal
scheme~\cite{Frixione:2008yi,ATL-PHYS-PUB-2016-020}.
Uncertainties due to the parton shower model are evaluated using samples interfaced to \HERWIG[7.04] with the default
parameters and the MMHT14LO PDF set.  Uncertainties in the matching of the parton shower to the ME are evaluated
using a sample generated with \MGNLO[2.6.2]~\cite{Alwall:2014hca} using the \NNPDF[2.3nlo] PDF set.

Diboson production ($VV$) with fully leptonic and semileptonic final states are modelled with the \SHERPA 2.2.11
generator using matrix elements calculated at NLO accuracy in QCD for up to one additional parton and at LO
accuracy for up to three additional parton emissions. The loop-induced $gg \rightarrow VV$ processes are modelled
using LO ME calculations for up to one additional parton emission.  The ME calculations
are matched to the \SHERPA parton shower using the \MEPSatNLO prescription. The virtual QCD corrections are provided
by the \OPENLOOPS library. The \NNPDF[3.0nnlo] set of PDFs is used, along with the dedicated set of tuned parton-shower
parameters developed by the \SHERPA authors.

Higgs boson production ($H \to \tau\tau$) via gluon--gluon fusion is simulated at next-to-next-to-leading-order (NNLO)
accuracy in QCD using \POWHEGBOX[v2]~\cite{Hamilton:2013fea,Hamilton:2015nsa,Alioli:2010xd,Nason:2004rx,Frixione:2007vw}.
The \PDFforLHC[15nnlo] PDF set~\cite{Butterworth:2015oua} and the \AZNLO tune of \PYTHIA[8] are used.  The
Monte Carlo sample is normalised to the next-to-NNLO cross-section in QCD plus
electroweak corrections at next-to-leading order
(NLO)~\cite{deFlorian:2016spz,Anastasiou:2016cez,Anastasiou:2015ema,Dulat:2018rbf,Harlander:2009mq,Harlander:2009bw,
Harlander:2009my,Pak:2009dg,Actis:2008ug,Actis:2008ts,Bonetti:2018ukf,Bonetti:2018ukf}.
The normalisation accounts for the decay branching ratio calculated with
HDECAY~\cite{Djouadi:1997yw,Spira:1997dg,Djouadi:2006bz} and
\PROPHECY~\cite{Bredenstein:2006ha,Bredenstein:2006rh,Bredenstein:2006nk}.

Potential off-shell leptoquark and $Z'$ boson contributions to high-mass Drell--Yan production are modelled with
\MGNLO[2.9.9] using LO-accurate matrix elements with up to two final-state partons.  The samples are generated
using the ‘vector\_LQ\_UFO' model~\cite{Baker:2019sli} and correspond separately to contributions from
BSM interference with the SM amplitude and from BSM-only amplitudes.  In the off-shell regime the kinematics of
$b$-jets are similar between SM and BSM amplitudes, so the calculation consists of an SM amplitude at LO with up
to two additional partons, reweighted to the predictions for several leptoquark and $Z'$ boson parameters.  The ME
calculation employs the \NNPDF[2.3lo] set of PDFs, and the events are interfaced to \PYTHIA[8.306]~\cite{Bierlich:2022pfr}
for the modelling of the parton shower, hadronisation, and underlying event. The A14 tune of \PYTHIA is used with the
\NNPDF[2.3lo] PDF set. The overlap between ME and parton shower emissions is removed using the
CKKW-L merging procedure.  The decays of bottom and charm hadrons are performed by \EVTGEN[1.2.0].

The \MGNLO calculation of Drell--Yan in association with $b$-jets is validated by comparing the ATLAS measurement of
$Z+b$-jet production~\cite{STDM-2018-43} to the predictions of \MGNLO generated at the same order and with the same
merging procedure as used for the signal sample.  The \MGNLO predictions agree with the measurement to within its
precision of 5\%.  The signal generation is further validated by comparing $\tau\tau b$ production using the leptoquark
signal model to the same process using the SM reweighted to the signal model.  The cross-sections agree to better than
10\%.  Uncertainties in the signal cross-sections from renormalization and factorization scale variations, and from PDF
variations, are of this order or larger.

Effective couplings involving $\tau$-leptons are modelled using \MGNLO[2.9.9] with ME calculations from the
SMEFTSim 3.0 ‘topU3l' model~\cite{Brivio:2020onw}.  Events are generated by setting one of the Wilson
coefficients to 1/TeV$^2$ and reweighting to the others.  In the fit to data the coefficients are rescaled as
necessary to correspond to those of the ‘top' model.  Beyond the ME-level the samples are processed
in the same way as for the leptoquark and $Z'$ boson samples.

Table~\ref{tab:mc} summarizes the modelled processes and the event generators used in the analysis.  For all samples, the
effect of multiple interactions in the same and neighbouring bunch crossings (\pileup) is modelled by overlaying the simulated
hard-scattering event with inelastic proton--proton (\(pp\)) events generated with \PYTHIA[8.186] using the \NNPDF[2.3lo]
set of parton distribution functions (PDF) and the A3 set of tuned parameters~\cite{ATL-PHYS-PUB-2016-017}.  The ATLAS
detector is modelled using a simulation~\cite{SOFT-2010-01} based on {\textsc{Geant4}}~\cite{Agostinelli:2002hh}.

\begin{table}[btp]
\centering
\caption{Monte Carlo samples used to model the SM and non-SM processes.  Additional samples are used to estimate
systematic uncertainties. }
\label{tab:mc}
{\renewcommand{\arraystretch}{1.3}
\begin{tabular}{l l l l }
\toprule
Process & Generator & Perturbative order (QCD) & PDF \\
\hline
Drell-Yan & \SHERPA & NLO ($\leq 2$ partons), & \NNPDF[3.0nnlo] \\
& & LO (3--5 partons) & \\
$t\bar{t}$, single top & \POWHEG + \PYTHIA8 & NLO & \NNPDF[3.0nlo] \\
$VV$ & \SHERPA & NLO ($\leq 1$ parton), & \NNPDF[3.0nnlo] \\
& & LO (2--3 partons) & \\
$H\to \tau\tau$ & \POWHEG + \PYTHIA8 & NNLO & \PDFforLHC[15nnlo] \\
Non-SM & \MGNLO & LO & \NNPDF[2.3lo] \\
& + \PYTHIA8 &  & \\
\bottomrule
\end{tabular}
}
\end{table}

\subsection{Misreconstructed \tauhad background}

Some \tauhad candidates originate from hadronic jets, i.e.\ they are misreconstructed.
Events with one jet misreconstructed as a $\tau_{\textrm{had}}$ object arise predominantly from
$W(\rightarrow \tau\nu)$+jets production, and events with two misreconstructed jets arise from QCD multijet
production. Both backgrounds are estimated using a `fake-factor' technique~\cite{ATLAS:2025akg}, where a
sample enriched in misreconstructed jets is used to predict the contribution in
the measurement sample.

The Tight $\tau_{\textrm{had}}$ objects used in this analysis include a requirement that the \RNN score be larger
than 0.8.  Exclusive background-enriched regions are selected by requiring one or two $\tau_{\textrm{had}}$ objects
to have $\RNN <0.8$.  The extrapolation to the selection region is performed using a fake factor, defined as the
ratio of misreconstructed jets satisfying $\RNN > 0.8$ to those with $\RNN < 0.8$.
The value of the fake factor depends on the parton-level source of the jet, and is $\approx 3$ times
larger for quark-initiated jets than for gluon-initiated jets.  The source of the jets in the background-enriched
region must therefore be estimated in order to obtain an accurate fake factor for extrapolation.

\begin{figure}[t!]
\centering
\subfloat[]{
\includegraphics[width=0.465\textwidth]{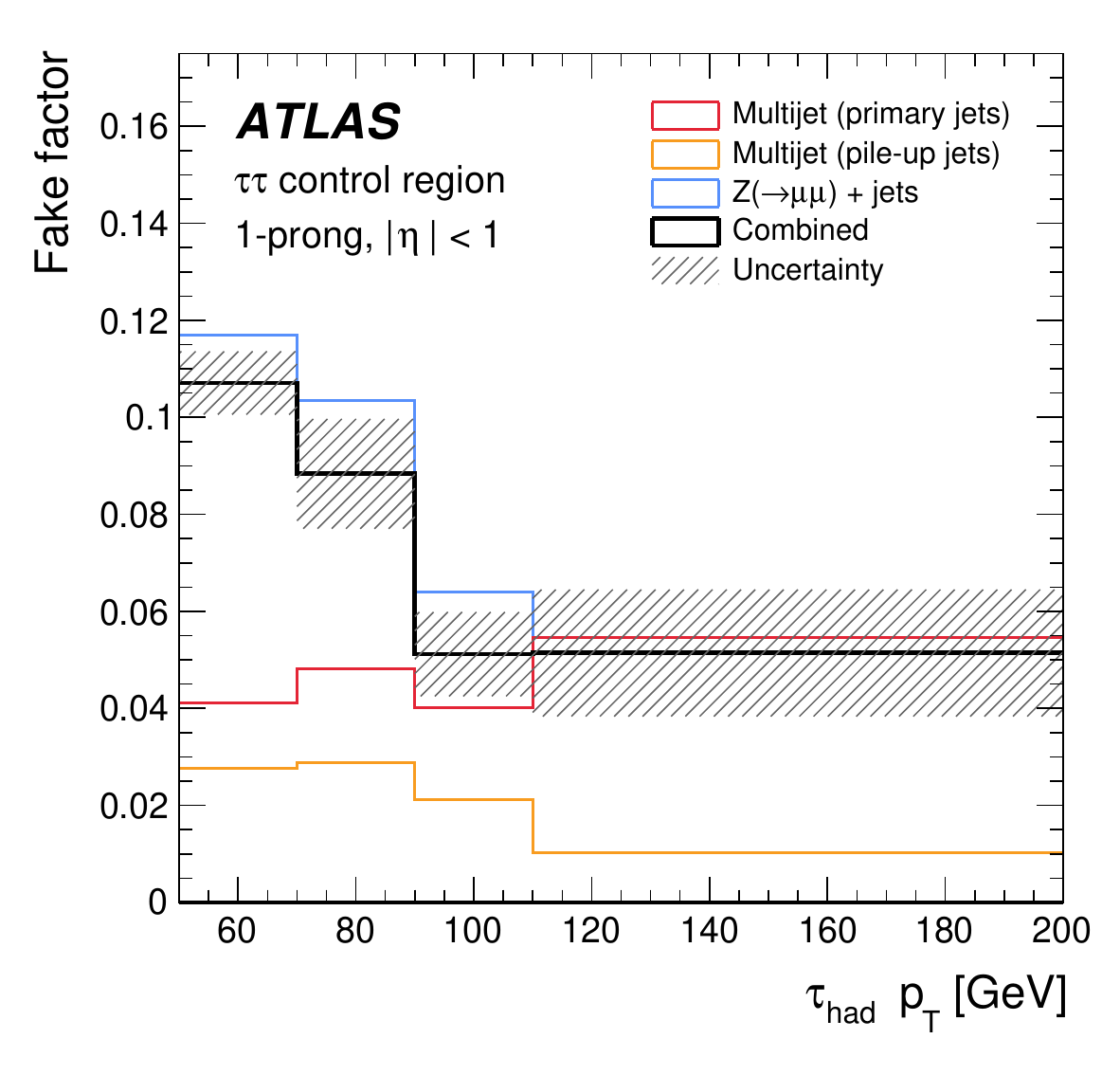}
}
\subfloat[]{
\includegraphics[width=0.525\textwidth]{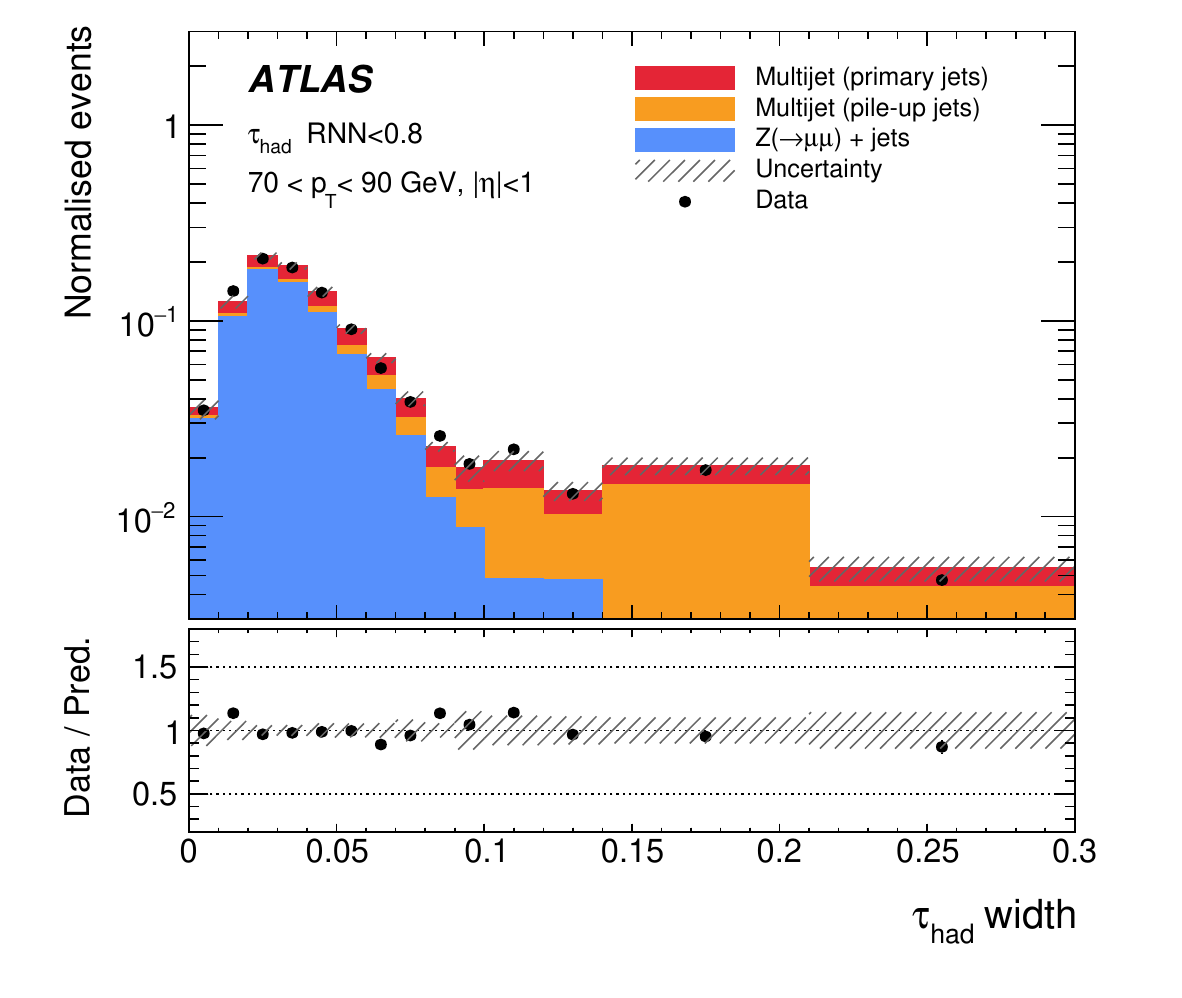}
}
\caption{(a) The fake factor as a function of $\tau_{\textrm{had}}$ \pt for each category of events and $\tau_{\textrm{had}}$
objects with one associated charged track and $|\eta|<1$.  The overall fake factor used in the analysis (black hatched) is
determined using the relative fractions of the categories obtained from the width fit. (b) The result of the width fit
for $\tau_{\textrm{had}}$ objects with one associated charged track, \pt between 70 and 90 GeV, and $|\eta|<1$.  The
ratio of data to prediction is shown in the bottom panel, along with the combined statistical uncertainty in the templates
(hatched band).  This fit is used to obtain the overall fake factor in the second bin of the left-hand plot. }
\label{fig:taufake}
\end{figure}

An effective discriminant for a jet's origin is its width, defined as the sum of the $\Delta R$-weighted \pt of its
constituents, divided by the total \pt~\cite{STDM-2011-38}.
The width of the $\tau_{\textrm{had}}$ object is calculated using its associated calorimeter clusters, and its distribution
is obtained in quark-enriched $Z(\rightarrow \mu\mu)$+jets events and in gluon-enriched multijet events~\cite{ATLAS:2025akg}.
Jets in the latter set of events are further divided into those arising from the hard scatter and those arising from pileup,
via the JVT algorithm.  The fake factor is calculated in each sample.  The fake factors are shown as a function of the \pt of
the $\tau_{\textrm{had}}$ object in Figure~\ref{fig:taufake}(a).

The normalization of each $\tau_{\textrm{had}}$ source in the background-enriched region is determined from
fits to the width distribution as a function of \pt and $\eta$, for one or three associated charged tracks.
Figure \ref{fig:taufake}(b) shows the results of a fit to the $\tau_{\textrm{had}}$ object width in events with
one $\tau_{\textrm{had}}$ object with $\RNN <0.8$ and one with $\RNN > 0.8$, where the object with $\RNN <0.8$
has one associated charged track, \pt between 70 and 90 GeV, and $|\eta|<1$.  Each fit determines the relative
amount of each jet source, and the extrapolation to the selection region is performed by applying the appropriate
source-weighted fake factor to each event.  The resulting yield correctly accounts for the $W$+jets misreconstruction
background, but it double-counts the contribution from the multijet background.  This overestimate is corrected by
subtracting this contribution, determined using events containing two $\tau_{\textrm{had}}$ objects with $\RNN <0.8$
multiplied by the appropriate fake factors.

\begin{figure}[t!]
\centering
\subfloat[]{
\includegraphics[width=0.49\textwidth]{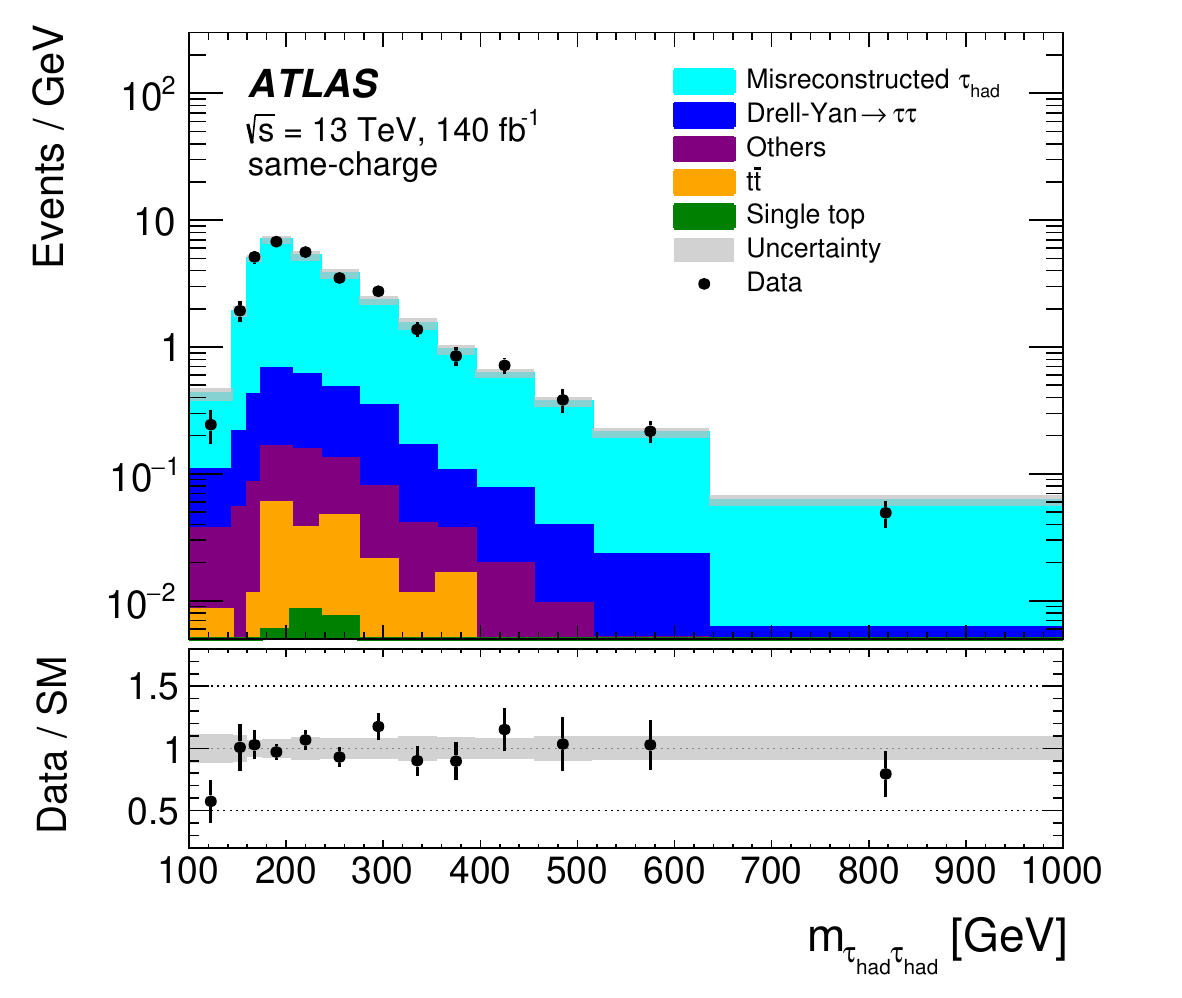}
}
\subfloat[]{
\includegraphics[width=0.49\textwidth]{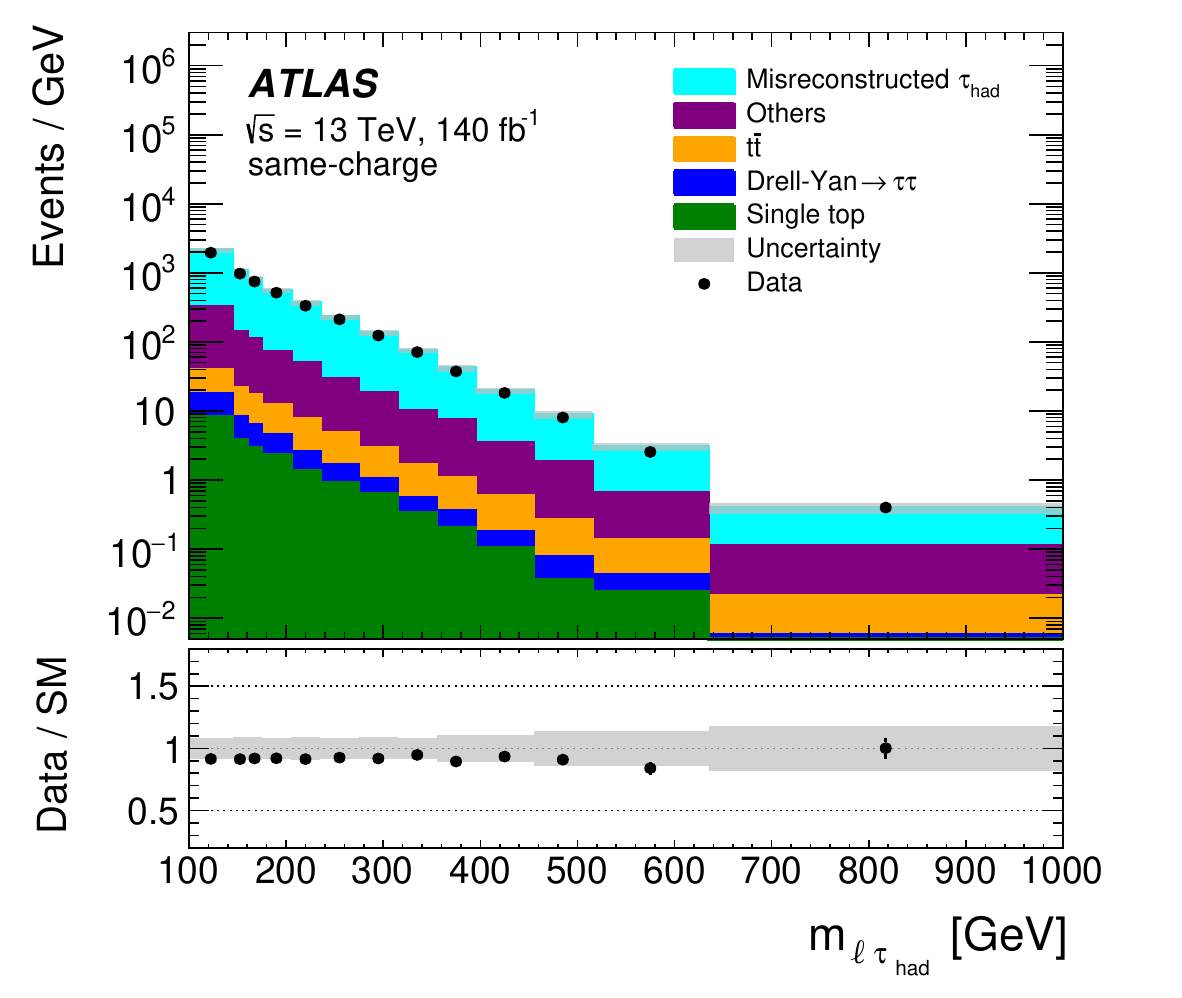}
}
\caption{The distribution of the visible invariant mass of (a) two same-charge \tauhad objects
or (b) one \tauhad object and an electron or muon with the same charge.  The bottom panels show
the ratio of data to the SM prediction, with the shaded band representing the total uncertainty
in the prediction.  The processes labelled `Others' are the production of dibosons, Higgs bosons,
and Drell--Yan pairs of electrons and muons.  The events are used to validate the estimate of the
misreconstructed \tauhad background.
}
\label{fig:taufakeval}
\end{figure}

The uncertainty in the background estimate is dominated by statistical uncertainties in the template histograms used
for the fit to the $\tau_{\textrm{had}}$ width, on the fit parameters, on the individual fake factors, and on the yield
of $\tau_{\textrm{had}}$ objects with $\RNN <0.8$.  The method is validated using events with two same-charge \tauhad
objects and events with one \tauhad object and a same-charge muon or electron.  The majority of these events contain a
misreconstructed \tauhad object.  Events with an electron also contain a sizeable contribution from the Drell--Yan
production of electron pairs, where an electron is misreconstructed as a $\tau_{\textrm{had}}$; this component is estimated
using simulation rather than included in the misreconstructed $\tau_{\textrm{had}}$ background.  Figure \ref{fig:taufakeval}
shows that the predictions are consistent with the data over a large range of reconstructed invariant mass ($m_{\hadhad}$ or
$m_{\lephad}$), given the approximately 10\% uncertainty in the prediction.

\section{Measurement of the inclusive \hadhad cross-section}
\label{sec:xsec}

The measured differential cross-section is defined by kinematic selections of final-state particles such that
they lie within a fiducial phase space which closely reflects the acceptance of the detector.
This ensures minimal extrapolation from the observed detector-level
data to the unfolded particle-level cross-sections, and minimal model
dependence in the final particle-level measurement.

\subsection{Definition of the measured final state}

Particle-level \tauhad candidates are built from the visible decay products of prompt, hadronically-decaying
\tauleptons; the momentum sum of these visible decay products (the `visible momentum') must yield
$\pt > 20$~\GeV\ and $\abseta < 2.47$
(excluding $1.37 < \abseta < 1.52$).
`Prompt' here means that the lepton does not have a hadron ancestor beyond the
initial-state protons.

Only events with exactly two fiducial \tauhads, as defined above, contribute to
the cross-section.
The \tauhad with the highest \pt is required to have $\pt > 90$~GeV, and the other is required to have $\pt > 60$~GeV.
They are required to have opposite signs, and their visible components are used to build the particle-level \mtautauvis.
As in Refs. \cite{STDM-2018-30} and \cite{STDM-2018-55}, fiducially accepted
events need not originate from any particular production process, be that
Drell--Yan, \ttbar, multi-boson, $tW$, or some other source. Any event that meets
the particle-level fiducial requirements is treated as signal and will be
included in the reported unfolded cross-sections.

\subsection{Correction for detector effects}

\begin{figure}[th!]
\centering
\includegraphics[width=0.75\textwidth]{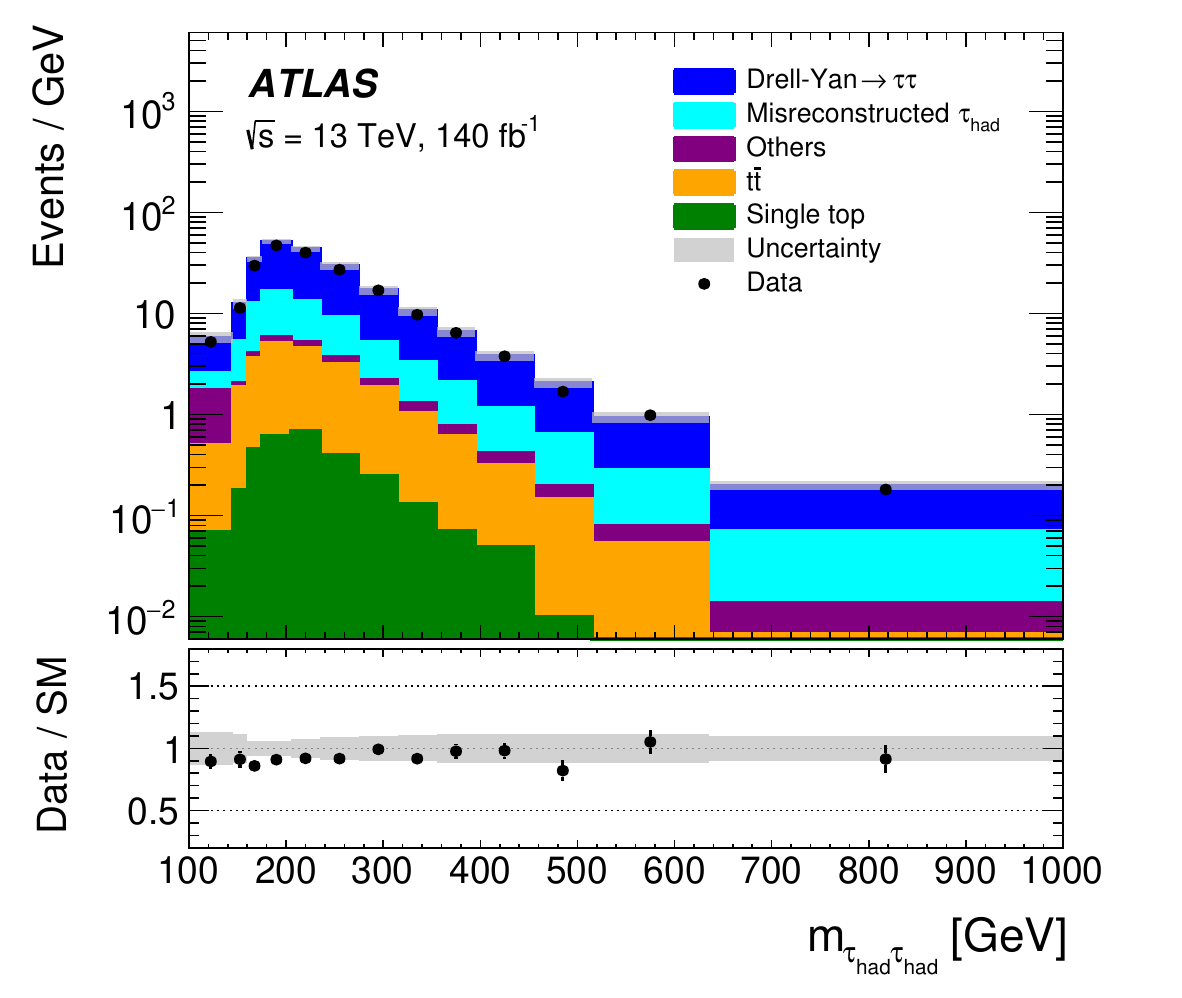}
\caption{The \tauhad\tauhad visible invariant mass in region of the measurement,
before background subtraction and correction for detector effects.
} \label{fig:detmtautauvis}
\end{figure}

The detector-level distribution of the visible \tauhad\tauhad invariant mass
is shown in Fig.~\ref{fig:detmtautauvis}. The detector-level data are
compared to the SM prediction consisting of the MC-estimated contributions from SM processes producing $\tau$ pairs,
as well as the contribution from misidentified \tauhads. The data are well modelled by the simulation.

Detector effects are unfolded using an iterative Bayesian
method~\cite{DAgostini:1994fjx} to yield a detector-agnostic measurement of
fiducial, differential cross-sections binned in \mtautauvis.
The key ingredients to such an unfolding are the reconstruction efficiency, the backgrounds from misidentified
\tauhads, and the migrations between particle and detector level caused by the detector response, as encoded in the transfer,
or migration, matrix. The method reweights the input distribution in a series of interations based on the level
of agreement between the data and the simulation, progressively removing the
bias from the input distributon.
Two iterations are used in this case, a number which was found to be the optimal
compromise between the bias, which decreases with each iteration, and the statistical uncertainty, which increases.

Non-fiducial events are defined as those that do not pass the particle-level
event selection: that is, they either fail the particle-level object selection
(for example they do not contain two true \tauleptons) or the object kinematics
fall out of the fiducial acceptance.
Non-fiducial events constitute around 27\% of the events in the bin around $\mtautauvis = 150$~GeV,
falling to around 4\% of the events at masses above 200~GeV.
Contributions from this background are taken from MC predictions and accounted
for as part of the unfolding procedure.

\begin{figure}[th!]
\centering
\includegraphics[width=0.75\textwidth]{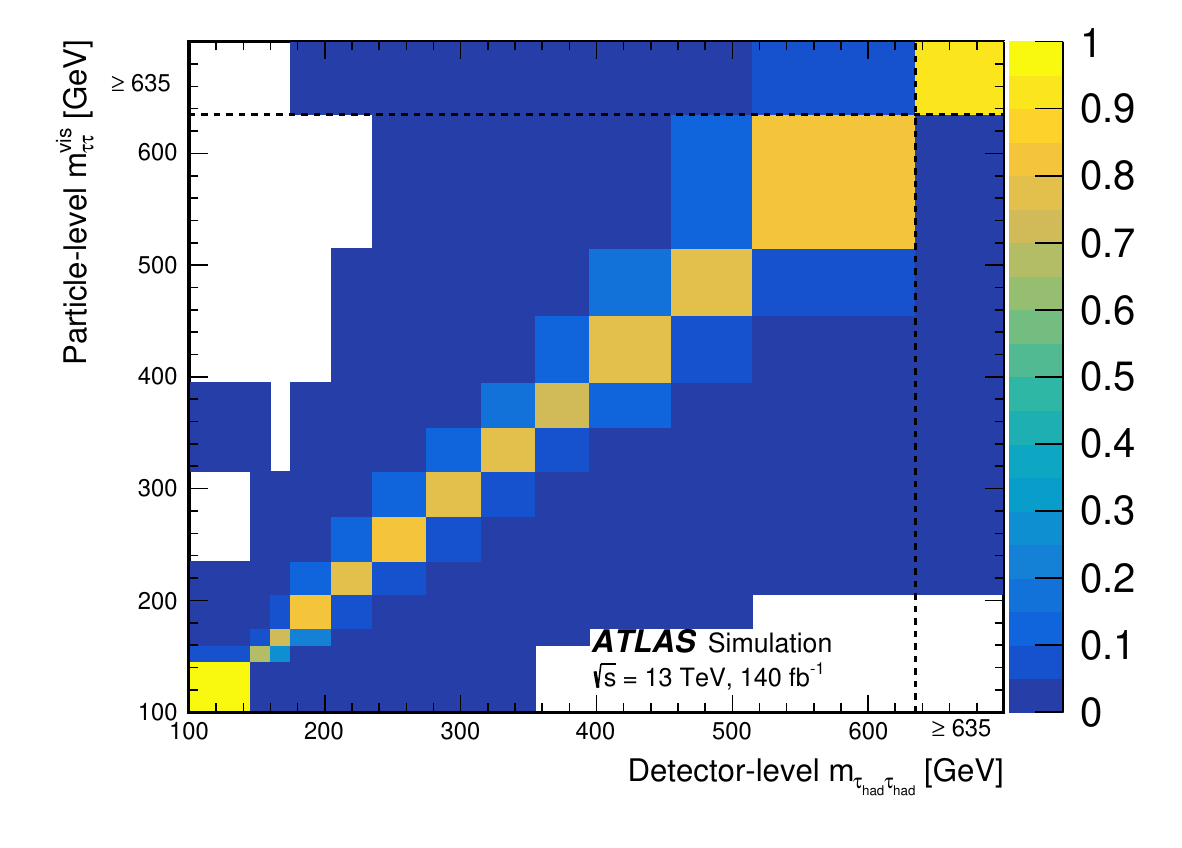}
\caption{The nominal transfer matrix, showing the probability of an event from a fiducial \mtautauvis
bin to be reconstructed at
detector-level in a given bin of \hadhad invariant mass.
} \label{fig:migration-matrix}
\end{figure}

For the fiducial events which also pass the event selection criteria at the detector level, there will be migrations between
mass bins due to the finite resolution of the detector. The transfer matrix describing these migrations using the off-diagonal
matrix cells is evaluated using the nominal \SHERPA samples, and is shown in Figure~\ref{fig:migration-matrix}.
The efficiency is the diagonal component of the migration matrix, divided by
the sum of the truth row for each bin plus the number of events with a true \mtautauvis in that bin but no reconstructed \taulepton pair.
It is around 15\% for $\mtautauvis > 200$~GeV.

The purity in each bin is the fraction of fiducial events in a given detector-level bin which are in the same bin at particle-level.
This is essentially the diagonal component of the migration matrix divided by the sum of the reconstruction
level column for each bin. It is about 70\% for masses of 150~GeV, and larger everywhere else.

The systematic uncertainties in the unfolded cross-section are evaluated by shifting the MC sample by the relevant uncertainty,
re-evaluating the transfer matrix and redoing the unfolding.
The total statistical uncertainty is evaluated similarly, by generating
bootstraps from a Poisson distribution~\cite{ATL-PHYS-PUB-2021-011}.
To assess any bias in the procedure, additional pseudodata sets were created by injecting simulated BSM events generated using the
leptoquark model discussed in the following section,
as well as by varying the SM subprocess composition of the sample (the most important being varying the contribution
of top processes compared to Drell--Yan within their uncertainties)
and by reweighting the visible \tauhad \pt distribution of the nominal SM
sample to data. In each case the (pseudo)data were then unfolded using the nominal response matrix derived from purely SM events. The biases were
found to be small, and where significant are included in the modelling uncertainty.

\begin{figure}[th!]
\centering
\includegraphics[width=0.75\textwidth]{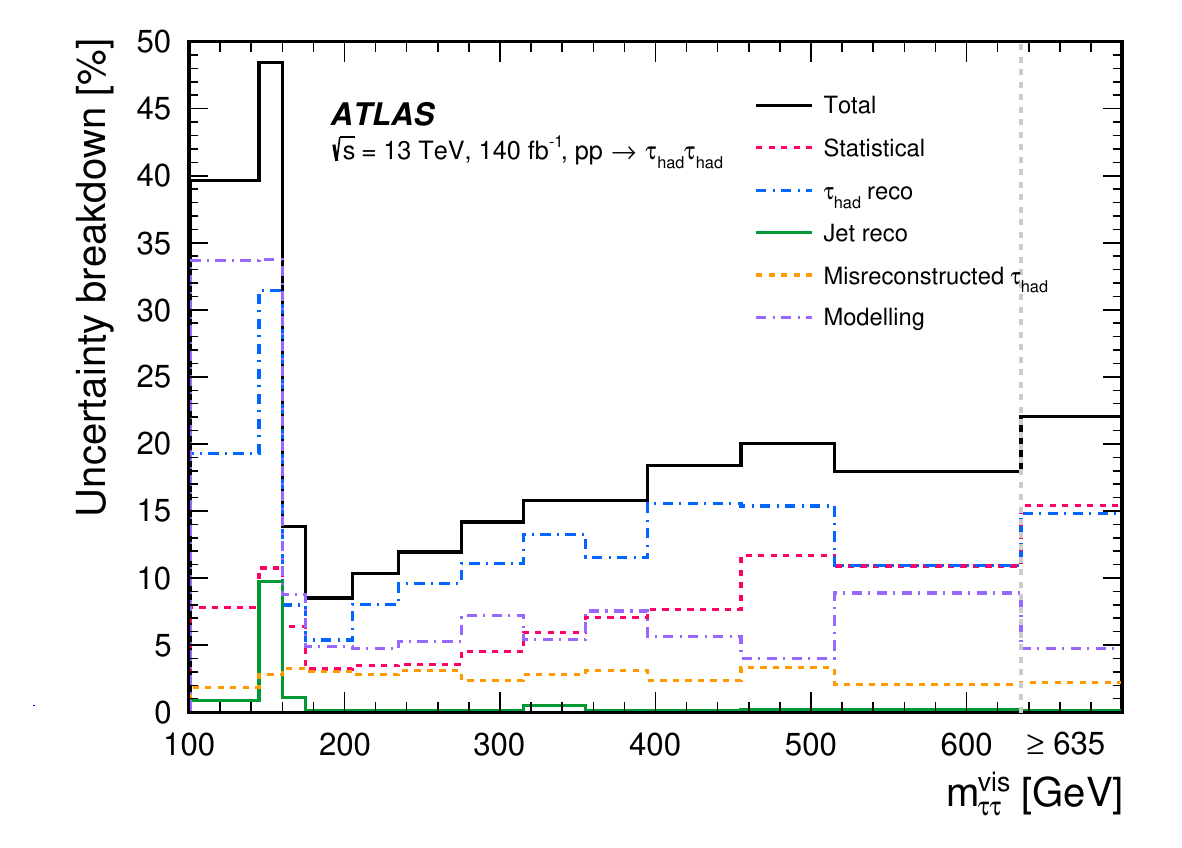}
\caption{Uncertainty breakdown for the \mtautauvis differential cross-section measurement.}
\label{fig:unfolding-error-breakdown}
\end{figure}

The systematic uncertainty breakdown obtained for each subsample is shown in Figure~\ref{fig:unfolding-error-breakdown}. It can be
seen that the most important uncertainties are the modelling and $\tau$-lepton identification uncertainties, especially near the low mass
threshold, with the statistical uncertainty rising from around 8\% at low $\mtautauvis$ to 15\% at the highest values.
The total uncertainty rises to close to 50\% at low mass, principally due to the fact that in this region
the $\tau$-lepton candidates lie close to the $\pt$ threshold.

\subsection{Results}
\label{sec:xsecresults}

\begin{figure}[htb]
\centering
\includegraphics[width=0.75\textwidth]{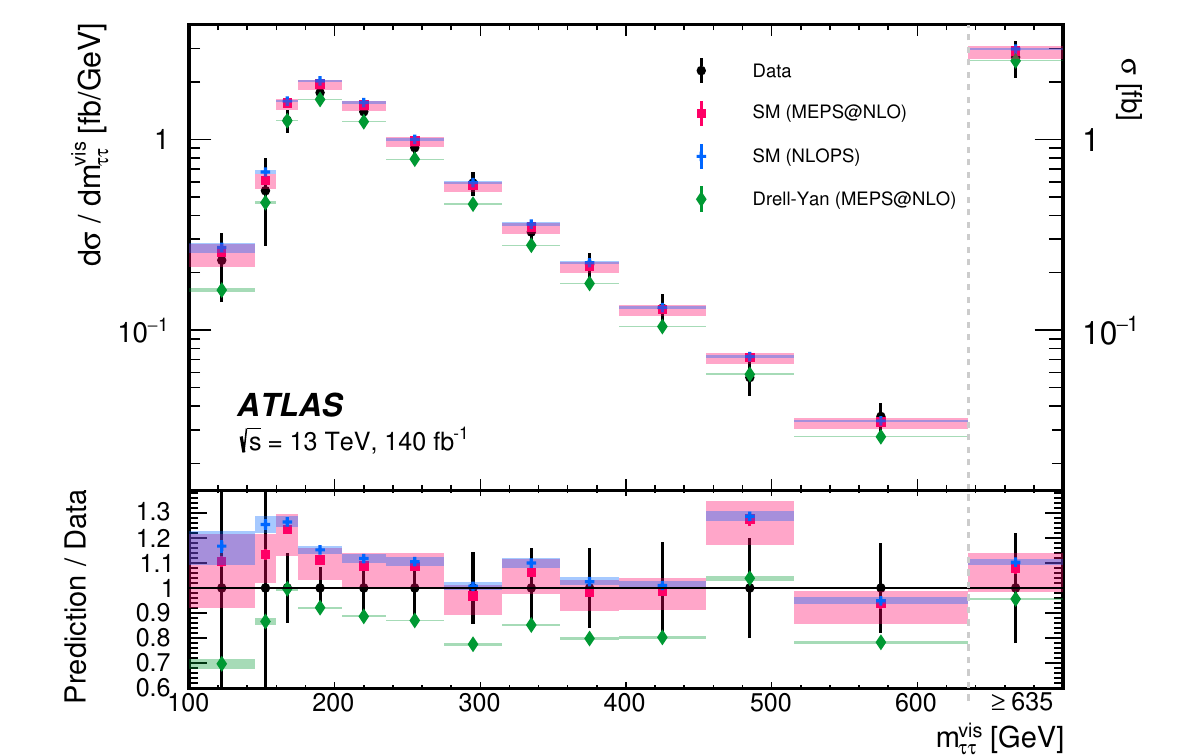}
\caption{Unfolded fiducial cross-section, differential in \mtautauvis, compared with SM predictions
from \MEPSatNLO (\SHERPA[2.2.11]) and NLOPS (\POWHEGBOX +\PYTHIA[8]),  where all SM processes that contribute significantly
to the cross-section are considered.
The curve labelled `Drell--Yan (\MEPSatNLO)' corresponds to \SHERPA[2.2.11]
generated Drell--Yan process only, shown with only MC statistical uncertainties,
to indicate the expected contribution from this process.
The rightmost bin is an overflow bin, integrating the cross-section above 635~GeV, as indicated on the right-hand axis
label.
}
\label{fig:thy_pred_unfold_xs}
\end{figure}

The measured cross-section is shown in Figure~\ref{fig:thy_pred_unfold_xs}.
It peaks at $\mtautauvis \approx 200$~GeV, falling in magnitude below this value
because only a small fraction of events at lower masses satisfy the transverse momentum
cuts of 90~GeV and 60~GeV which form part of the definition of the fiducial cross section.
The measurement is compared to SM predictions from \SHERPA[2.2.11] (\MEPSatNLO) and \POWHEGBOX +\PYTHIA[8] (NLOPS), where all
SM processes that contribute significantly to the cross-section are considered (principally top production and Drell--Yan).
The uncertainty bars on the data contain all experimental uncertainties. The uncertainties in the SM predictions
include the effect of QCD renormalization and factorization scale variations, and PDF uncertainties.

The theory predictions from both \MEPSatNLO and NLOPS lie within the uncertainty bars of the unfolded data.
At lower masses the predictions lie somewhat above the data, with the agreement improving for $\mtautauvis > 300$~GeV.
The covariance matrix for the uncertainties is constructed, under the assumption
that the theory and experimental uncertainties are normally distributed. Using this matrix, the p-value for the fit is 0.78.
The results are made available in HEPData~\cite{Maguire:2017ypu}, with an accompanying Rivet routine~\cite{Bierlich:2019rhm}.

\section{Search for new non-resonant interactions}
\label{sec:search}
%

Generic non-resonant interactions can be described by an effective field theory (EFT).
Of particular interest in the $\tau\bar{\tau}$ final state are preferential couplings to third-generation
fermions.  Such couplings would enhance the effect on the $\tau\bar{\tau}$ mass distribution in events
with one or two $b$-jets, as the presence of $b$-jets suggests the primary interaction involves $b$-quarks.
Since the background also depends on $b$-jet multiplicity, the analysis is separated into events with 0, 1,
or at least 2 $b$-jets.  A profile likelihood fit to the data is performed for general $\tau\tau qq$ and
$Z(/\gamma) \tau\tau$ EFT couplings, and for leptoquarks and $Z'$ bosons coupling preferentially to
$b$-quarks.

\subsection{SM effective field theory}
\label{sec:SMEFT}
The general Lagrangrian of an EFT respecting the Standard Model gauge symmetries can be expressed as:
\begin{equation}{\cal{L}}_{\textrm{SMEFT}} = {\cal{L}}_{\textrm{SM}} + \sum_i \frac{c_i {\cal{O}}^{(6)}_i}{\Lambda^2} + ...,
\end{equation}

\noindent
where the operators ${\cal{O}}^{(6)}$ have mass-dimension six, the suppression mass scale is $\Lambda$, $c_i$
are dimensionless coupling coefficients, and mass-dimension-five operators are neglected since they are
lepton-number-violating and strongly constrained by neutrino measurements~\cite{PhysRevD.103.115014}.  The
EFT assumes that the new-physics scale is above the mass scale directly accessible to the measurement.

Assuming no flavour symmetries and no baryon-number-violating operators, there are 2499 independent operators in the
Standard Model effective field theory (SMEFT).  To reduce the parameter space, the ‘top' model assumes
the first two quark generations are flavour-symmetric.  The coupling coefficients of this model have been constrained
by prior ATLAS global fits to measured observables~\cite{HIGG-2022-17}.  The operators relevant to $pp \to \tau\bar{\tau}$
production are four-fermion operators, operators with $V\tau\tau$ interactions, and the four-lepton operator that affects
muon decay, since the muon lifetime is used to set the weak coupling $G_F$.

The probed ‘top' operators are shown in Table~\ref{tab:CoeffOperators}.  Four-fermion operators involving $b$-quarks
are not considered since the leptoquark and $Z'$-boson models map onto these operators.  Operators connecting a
left-handed fermion to its right-handed counterpart lead to cross-sections too small to be probed by this
analysis.  The coefficient defined as $c_{\tau \gamma}$ is a linear combination of ‘top' operators that
contributes to the anomalous magnetic moment of the $\tau$-lepton, \atau, and to the electric dipole moment \dtau
if $c_{\tau \gamma}$ is complex~\cite{ESCRIBANO1993419,Beresford:2019gww,Beresford:2024dsc,Haisch:2023upo}:
\begin{equation*}
\delta \atau = \frac{2 \sqrt{2} m_\tau v}{e}\frac{\text{Re}[c_{\tau\gamma}]}{\Lambda^2},
\qquad
\delta \dtau = \sqrt{2} v \frac{\text{Im}[c_{\tau\gamma}]}{\Lambda^2},
\end{equation*}
where $v$ is the Higgs field vacuum expectation value, $e$ is the electron electric charge, $m_\tau$ is the \taulepton
mass, and $\delta$ indicates deviations from the Standard Model anomalous moments.  The corresponding coefficient for
$Z\tau\tau$ interactions is defined as $c_{\tau Z}$.  While the coefficients are taken to be real in the event generation,
constraints can nonetheless be inferred for $\delta \dtau$ (see Section~\ref{sec:results}).

The scale $\Lambda$ is assumed to be sufficiently high that the interference between the SMEFT and SM diagrams
dominates.  Fits performed to individual coupling coefficients using this interference as the signal are referred to
as `linear' fits since they depend linearly on the coupling coefficients.  As a check of the validity of the constraints,
an additional fit is performed including both the interference and the pure SMEFT contribution for each coefficient
assuming $\Lambda = 1$ TeV.  These fits are referred to as `linear plus quadratic' fits.  In the case of the
operators contributing to $c_{\tau\gamma}$ and $c_{\tau Z}$, the interference is suppressed by $m_\tau/\Lambda$
and the dominant contribution is from the pure-SMEFT quadratic term.

\begin{table}[btp]
\centering
\caption{Coupling coefficients $c_i$ and corresponding SMEFT operators $\mathcal{O}^{(6)}_i$ used in this analysis.
The operator and coefficient notation follows the convention of the SMEFTSim 3.0 ‘top' model, %
with $e$ replaced by $\tau$ to explicitly indicate the probed lepton flavour.  Right-handed states are labelled
by particle type, with $u$ referring to an up or charm quark and $d$ referring to a down or strange quark.
Left-handed states are labelled $q$ for the quark doublet of either of the first two generations and $l$, $\mu$, or $e$
for the $\tau$-lepton, muon, or electron doublet.  The cosine and sine of the Weinberg angle are denoted $c_W$ and $s_W$,
respectively.  The coupling coefficients are assumed to be real, and any necessary Hermitian conjugates are implied. }
\label{tab:CoeffOperators}
{\renewcommand{\arraystretch}{1.3}
\begin{tabular}{c c}
\toprule
Coupling coefficient & Operator \\
\hline
$c_{lq}^{(3)}$ & \scalebox{0.8}{$(\bar l \sigma^i \gamma^\mu l)(\bar q \sigma^i \gamma^\mu q)$} \\ \interrowfill
$c_{lq}^{(1)}$ & \scalebox{0.8}{$(\bar l \gamma^\mu l)(\bar q \gamma^\mu q)$} \\ \interrowfill
$c_{l u}$ & \scalebox{0.8}{$(\bar l \gamma^\mu l)(\bar u \gamma^\mu u)$} \\ \interrowfill
$c_{l d}$ & \scalebox{0.8}{$(\bar l \gamma^\mu l)(\bar d \gamma^\mu d)$} \\ \interrowfill
$c_{q \tau}$ & \scalebox{0.8}{$(\bar \tau \gamma^\mu \tau)(\bar q \gamma^\mu q)$} \\ \interrowfill
$c_{\tau u}$ & \scalebox{0.8}{$(\bar \tau \gamma^\mu \tau)(\bar u \gamma^\mu u)$} \\ \interrowfill
$c_{\tau d}$ & \scalebox{0.8}{$(\bar \tau \gamma^\mu \tau)(\bar d \gamma^\mu d)$} \\ \interrowfill
\hline
$c_{l l}$ & \scalebox{0.8}{$(\bar \mu \gamma^\mu \mu)(\bar e \gamma^\mu e)$} \\ \interrowfill
\hline
$c_{Hl}^{(3)}$ & \scalebox{0.8}{$(H^\dag i\overleftrightarrow{D}^i_\mu H)(\bar l \sigma^i \gamma^\mu l)$} \\ \interrowfill
$c_{Hl}^{(1)}$ & \scalebox{0.8}{$(H^\dag i\overleftrightarrow{D}_\mu H)(\bar l \gamma^\mu l)$} \\ \interrowfill
$c_{H\tau}$ & \scalebox{0.8}{$(H^\dag i\overleftrightarrow{D}_\mu H)(\bar{\tau} \gamma^\mu \tau)$} \\ \interrowfill
\hline
$c_{\tau W}$ & \scalebox{0.8}{$(\bar{l} \sigma^{\mu\nu}\tau)\sigma^i H W^i_{\mu\nu}$} \\ \interrowfill
$c_{\tau B}$ & \scalebox{0.8}{$(\bar{l} \sigma^{\mu\nu}\tau)H B_{\mu\nu}$} \\ \interrowfill
$c_{\tau Z}$ & \scalebox{0.8}{$(\bar{l} \sigma^{\mu\nu}\tau)\sigma^i H (c_W \sigma^i W^i_{\mu\nu} + s_W B_{\mu\nu})$} \\ \interrowfill
$c_{\tau \gamma}$ & \scalebox{0.8}{$(\bar{l} \sigma^{\mu\nu}\tau)\sigma^i H (-s_W \sigma^i W^i_{\mu\nu} + c_W B_{\mu\nu})$} \\ \interrowfill
\bottomrule
\end{tabular}
}
\end{table}

\subsection{Leptoquark models}

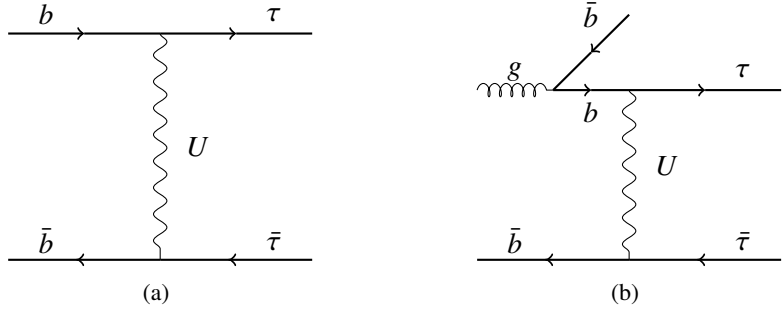
\begin{figure}[t!]
\centering
\subfloat[]{
\ignore{ \includegraphics[width=0.49\textwidth]{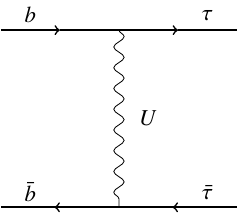} }

\begin{tikzpicture}

\draw[thick,->] (1.,4.75) --(2.,4.75);
\draw[thick,-] (2.,4.75) --(3.,4.75);
\draw[thick,->] (3.,4.75) --(4.,4.75);
\draw[thick,-] (4.,4.75) --(5.,4.75);
\draw[thick,-<] (1.,1.75) --(2.,1.75);
\draw[thick,-] (1.9,1.75) --(3.,1.75);
\draw[thick,-<] (3.,1.75) --(4.,1.75);
\draw[thick,-] (3.9,1.75) --(5.,1.75);
\draw[thick] (1.5,5) node {$b$};
\draw[thick] (4.5,5) node {$\tau$};
\draw[thick] (1.5,2.) node {$\bar{b}$};
\draw[thick] (4.5,2.) node {$\bar{\tau}$};

\draw[decorate,decoration=snake,-]  (3.,4.75) -- (3.,1.75) node [anchor=east]{};
\draw[thick] (3.5,3.25) node {$U$};

\end{tikzpicture}
}
\hspace{50 pt}
\subfloat[]{
\ignore{ \includegraphics[width=0.49\textwidth]{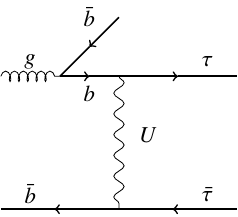} }

\begin{tikzpicture}
\draw[decorate,decoration=snake,-]  (8.,4.) -- (8.,1.75) node [anchor=east]{};
\draw[thick] (8.5,3.) node {$U$};
\draw[decorate,decoration=coil, segment length=1.5mm]  (6.,4.) -- (7.,4.) node [anchor=east]{};
\draw[thick,->] (8.,5.) --(7.5,4.5);
\draw[thick,-] (7.,4.) --(7.5,4.5);
\draw[thick,->] (7.,4.) --(7.5,4.);
\draw[thick,->] (7.5,4.) --(9.,4.);
\draw[thick,-] (9.,4.) --(10.,4.);
\draw[thick,-<] (6.,1.75) --(7.,1.75);
\draw[thick,-] (6.9,1.75) --(8.,1.75);
\draw[thick,-<] (8.,1.75) --(9.,1.75);
\draw[thick,-] (8.9,1.75) --(10.,1.75);
\draw[thick] (7.5,5) node {$\bar{b}$};
\draw[thick] (7.5,3.7) node {$b$};
\draw[thick] (9.5,4.25) node {$\tau$};
\draw[thick] (6.5,2.) node {$\bar{b}$};
\draw[thick] (9.5,2.) node {$\bar{\tau}$};
\draw[thick] (6.5,4.25) node {$g$};
\end{tikzpicture}
}
\caption{The leading non-resonant diagrams for an initial state of (a) $b\bar{b}$ or (b) $g\bar{b}$ and a final
state of (a) $\tau\bar{\tau}$ or (b) $\tau\bar{\tau}\bar{b}$.  }
\label{fig:lqfeyn}
\end{figure}

The Lagrangian terms relevant for non-resonant $\tau\bar{\tau}$ production from t-channel vector leptoquark ($U$) exchange
are~\cite{Baker:2019sli}:

\begin{equation}
\Delta{\cal{L}} = U^\mu(\beta^{i3}_L \bar{q}^i_L \gamma_\mu \ell_L^3 + \beta^{i3}_R \bar{q}^i_R \gamma_\mu \tau_R)/\sqrt{2} + h.c.,
\end{equation}

\noindent
where $\beta^{i3}_L$ and $\beta^{i3}_R$ are the couplings to the left-handed and right-handed fermion states, respectively,
$q^i_L$ is the left-handed quark doublet of generation $i$, and $\ell_L^3$ is the third-generation left-handed lepton doublet.
For $i=3$, the dominant non-resonant diagrams consist of either an initial $b\bar{b}$ pair or an initial state of a gluon and
a $b$ or $\bar{b}$, where the gluon splits into a $b\bar{b}$ pair (Figure~\ref{fig:lqfeyn}).  A leptoquark is exchanged in the
t-channel to produce a pair of $\tau$-leptons.

The signal diagrams destructively interfere with the dominant SM process of Drell--Yan production.  For values of
$\beta^{i3}_{L,R}$ larger than one, there can be large cancellations between the interference and the pure-BSM contribution.
However, the two contributions have different $\tau\tau$ mass dependence, which can be exploited to improve sensitivity.  The
maximum sensitivity to the process comes from events with one associated $b$-jet.  For such events within a range of visible mass
of 100--250 GeV, interference contributes $-34 \pm 4$ events while pure-BSM contributes $19 \pm 3$ events for a leptoquark mass
of 2 TeV and $(\beta^{33}_L)^2 = 10$; in the mass range 400-1000 GeV the contributions are $-11 \pm 2$ and $39 \pm 12$ events,
respectively, where the uncertainties are statistical.

The signal samples are generated with \MGNLO, as described in Section~\ref{sec:modelling}.  Events are generated using
the SM process and subsequently reweighted to the signal process for a variety of coupling and leptoquark-mass ($m_U$)
values.  Other values are obtained by interpolating according to the expected non-resonant scaling,
$\sigma \propto (\beta^{i3}_{L,R})^2/m_U^2$ for interference
and $\sigma \propto (\beta^{i3}_{L,R})^4/m_U^4$ for the pure-BSM contribution.  Weights are included for the
scenarios preferred by the $B$-hadron anomalies, $\beta_R^{33} = -\beta_L^{33}$ and $\beta_L^{33} \neq 0$, with
other couplings equal to zero.  Additional weights correspond to scenarios of pure right-handed couplings,
$\beta_R^{33} \neq 0$, and left-handed couplings between second-generation quarks and third-generation leptons,
$\beta_L^{23} \neq 0$.

Scalar leptoquarks are also probed according to the Lagrangian \cite{Dorsner:2016wpm}:

\begin{equation}
\Delta{\cal{L}} = \lambda \bar{b}^C_R \tilde{S}_1 \tau_R + h.c.,
\end{equation}

\noindent
where $C$ stands for charge conjugation and $\tilde{S}_1$ represents the leptoquark field.  Signal samples are generated in
the same manner as for vector leptoquarks, i.e. reweighting the SM process to that of scalar leptoquark exchange for a
variety of mass values.

\subsection{$Z'$ boson model}

The Lagrangian terms relevant for $\tau\bar{\tau}$ production from s-channel $Z'$ boson exchange via $b\bar{b}$ annihilation
are~\cite{Baker:2019sli}:

\begin{equation}
\Delta{\cal{L}} = Z'^\mu(\zeta_q \bar{q}^3_L \gamma_\mu q^3_L + \zeta_u \bar{t}_R \gamma_\mu t_R +
\zeta_d \bar{b}_R \gamma_\mu b_R - 3\zeta_\ell \bar{\ell}_L^3 \gamma_\mu \ell_L^3 -
3 \zeta_\tau \bar{\tau}_R \gamma_\mu \tau_R)/(2\sqrt{6}),
\end{equation}

\noindent
where $\zeta_q$ ($\zeta_\ell$) and $\zeta_d$ ($\zeta_\tau$) are the couplings to third-generation left-handed and
right-handed quark (lepton) states, respectively.  The interference between the $Z'$ boson diagram and SM $\tau\bar{\tau}$
production is constructive.  The signal is modelled using the same \MGNLO samples as for leptoquark production.
Weights are included for the benchmark cases where the $Z'$ boson couples universally to left-handed fermions
($\zeta_q = \zeta_\ell \neq 0$) or to right-handed fermions ($\zeta_d = \zeta_\tau \neq 0$).  Non-resonant production
is ensured by reweighting to a $Z'$ boson mass of 3 TeV and extrapolating to other mass points by approximating the
propagator as the inverse square of the mass.

\subsection{Fit frameworks}

The various models are fit either to the observed data yields using a profile likelihood fit, or to the particle-level
differential cross-section (Figure~\ref{fig:thy_pred_unfold_xs}) using a $\chi^2$ minimization.

The profile likelihood fit is performed using the yields in three bins of \bjet multiplicity (0, 1, $\geq 2$) and
three bins of invariant mass of the \hadhad system (100--250, 250--400, and 400--1000 GeV).  The bins are chosen to
give the optimal sensitivity given the Monte Carlo statistical uncertainty of the signal sample.  The parameter of
interest is the square of the coupling tested in the model.  Systematic uncertainties are included as nuisance
parameters with Gaussian constraint terms in the likelihood function.  The normalizations of the Drell--Yan and
$t\bar{t}$ processes are freely floated in the fit, and are largely constrained by events with \bjet multiplicities
of 0 and $\geq 2$, respectively.

Systematic uncertainties are fully correlated across the three bins, with two exceptions.  The generator modelling
uncertainty in the Drell--Yan process is uncorrelated across \bjet multiplicity since the underlying source of
uncertainty is expected to differ for each multiplicity value.  The misreconstructed \tauhad background uncertainty
is also uncorrelated across \bjet multiplicity since separate data sets are used in the extrapolation from low values
to high values of the RNN discriminant in the background estimation.

The likelihood fit uses the {\texttt{pyhf}}~\cite{pyhf_joss} framework, which finds the shape of the likelihood function
around the maximum.  The {\texttt{cabinetry}}~\cite{refId0} toolkit is used to set up the fit and provide information on
nuisance parameter impacts, pulls, and correlations.  Exclusion limits are determined using a modified CLs
technique~\cite{Junk:1999kv,Read:2002hq} with the profile-likelihood ratio as the test statistic, using the
asymptotic approximation~\cite{Cowan:2010js}.  The fit model is constructed such that the parameter of interest scales
linearly with the interference term and quadratically with the pure-BSM term.

The $\chi^2$ fit is only performed for the SMEFT operator coefficients, which affect inclusive Drell--Yan production.
The $\chi^2$ is evaluated as described in Section~\ref{sec:xsecresults}, with the addition of a signal term
depending upon the SMEFT coupling and the inclusion of the corresponding signal uncertainties in the covariance matrix.
The best-fit coupling is evaluated, and the likelihood ratio with respect to the likelihood at this value
is used to evaluate a confidence level (CL) as a function of coupling.  The general procedure is as described in detail in
Ref.~\cite{Buckley:2021neu}\footnote{In the previous versions of \textsc{Contur}~\cite{Butterworth:2016sqg,Buckley:2021neu},
no signal fit was performed.  The current version, Contur~3.1, uses \textsc{Spey}~\cite{Araz:2023bwx} to obtain the
best-fit signal.}.

\subsection{Results}
\label{sec:results}

\begin{figure}[t!]
\centering
\subfloat[]{
\includegraphics[width=0.49\textwidth]{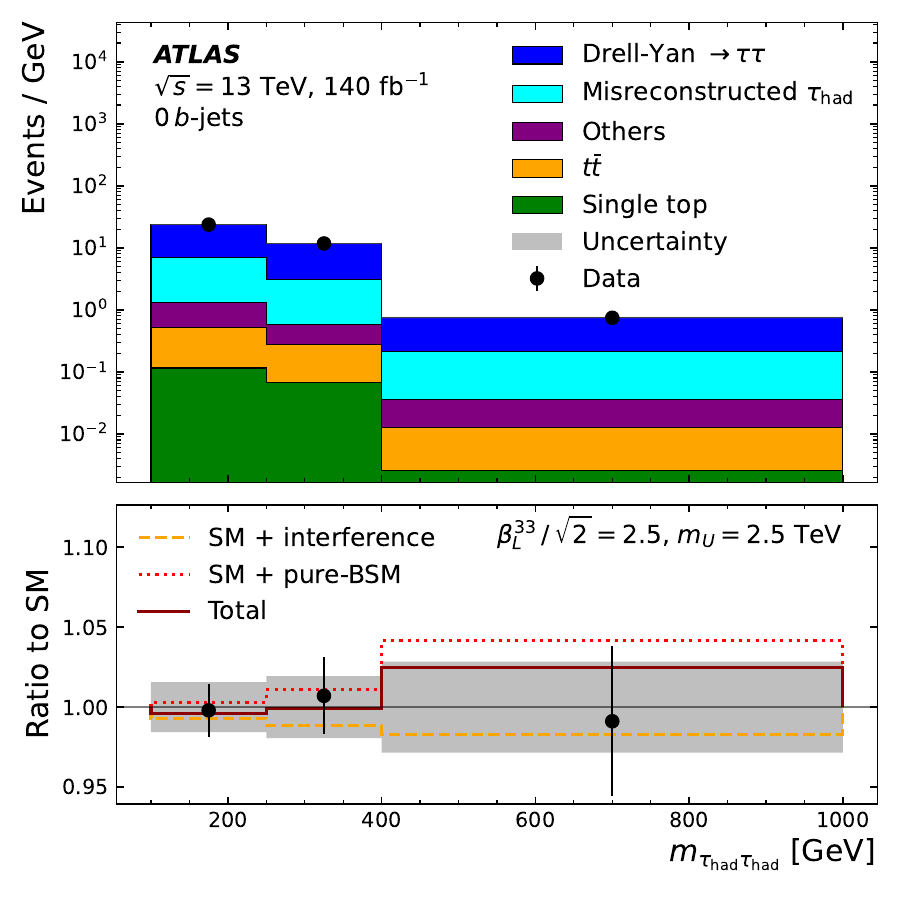}
}
\subfloat[]{
\includegraphics[width=0.49\textwidth]{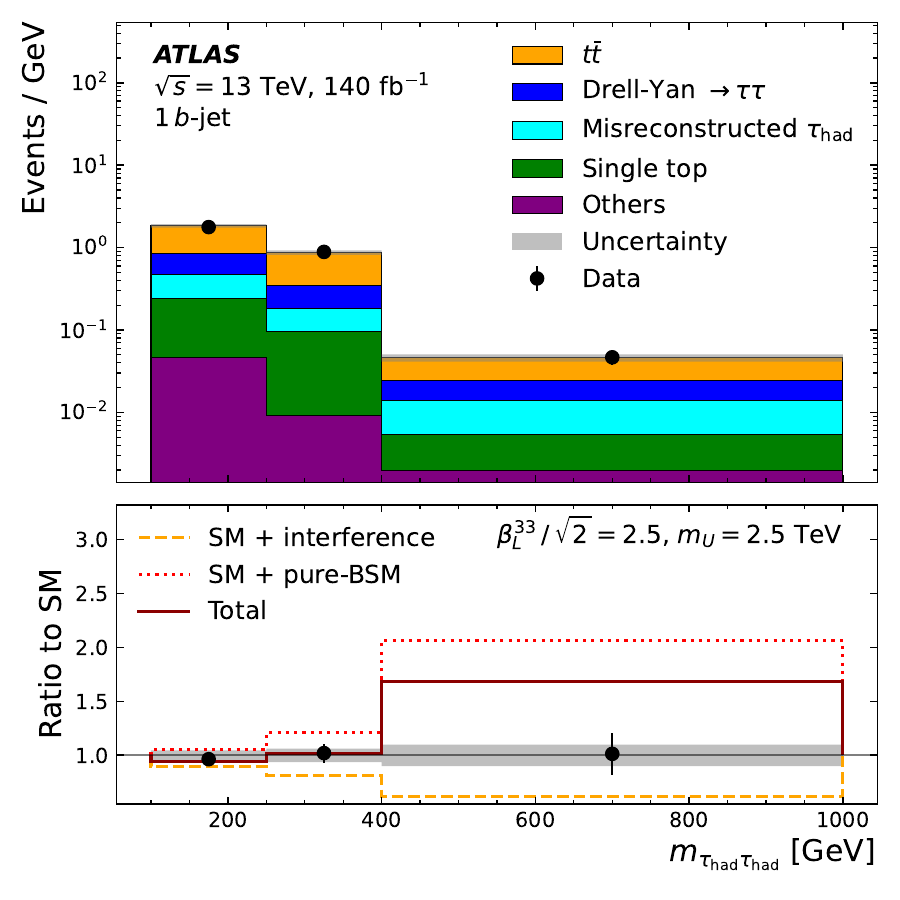}
} \\
\subfloat[]{
\includegraphics[width=0.49\textwidth]{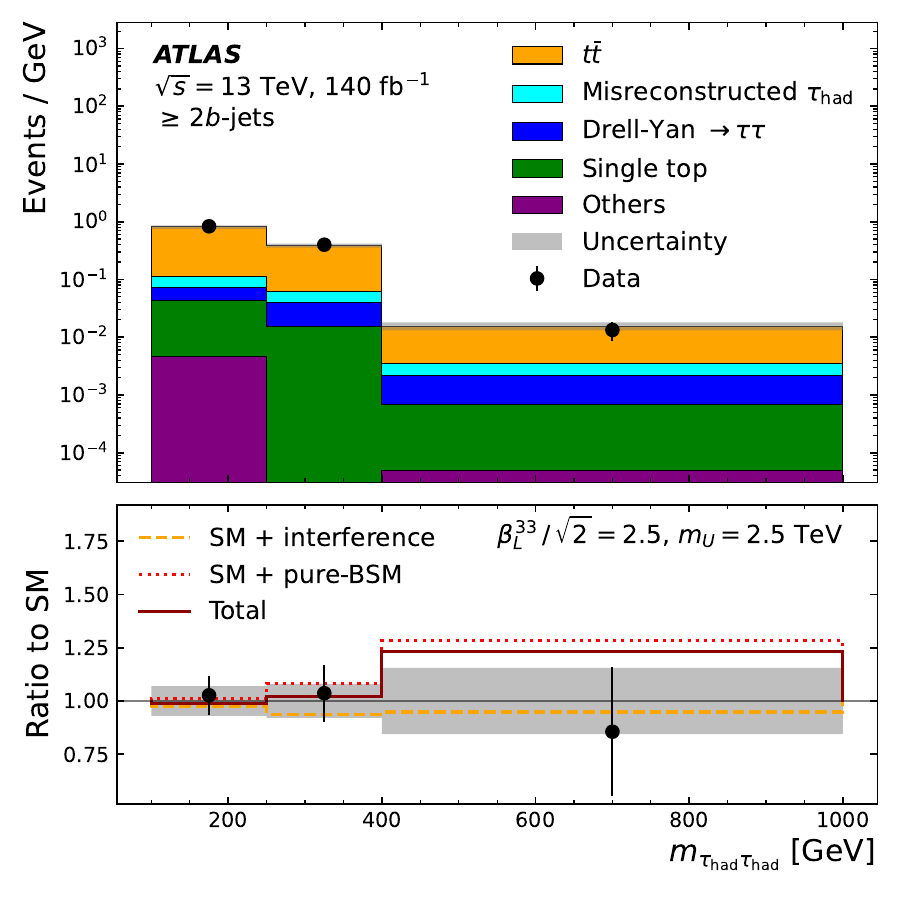}
}
\caption{ The background-only post-fit \hadhad visible invariant mass in events with (a) zero $b$-jets, (b) one $b$-jet,
or (c) at least two $b$-jets.
Shown are the data and the individual SM backgrounds, along with the ratio
of the data to the SM.
Also shown are the ratios of the sum of the SM prediction and either the
interference, pure-BSM, or total signal to the SM prediction, for the case
$\beta_L^{33}/\sqrt{2} = 2.5$ and a vector leptoquark mass of 2.5 TeV.  The
grey band shows the relative post-fit uncertainty on the SM prediction.
}
\label{fig:bjetmult}
\end{figure}

Figure~\ref{fig:bjetmult} compares the $m_{\hadhad}$ data distribution to the result of the SM-only fit and its
sum with the $m_{\textrm{LQ}} = 2.5$~TeV leptoquark signal model with a coupling value of $\beta_L^{33}/\sqrt{2} = 2.5$
that is excluded by this measurement at the 95\% CL.  The difference between the model and the data is less than
one standard deviation in each bin of \bjet multiplicity.  The fit gives overall multiplicative normalization
factors of $0.88 \pm 0.07$ and $0.80 \pm 0.10$ for the Drell--Yan and $t\bar{t}$ processes, respectively.  The
leading uncertainties are the Drell--Yan modelling uncertainties in events with one $b$-jet, and the
$t\bar{t}$ normalization and modelling uncertainties.  These are followed by the MC statistical uncertainties
in the bin with the highest $m_{\hadhad}$ in $1b$ events.

\begin{figure}[t!]
\centering
\includegraphics[width=0.95\textwidth]{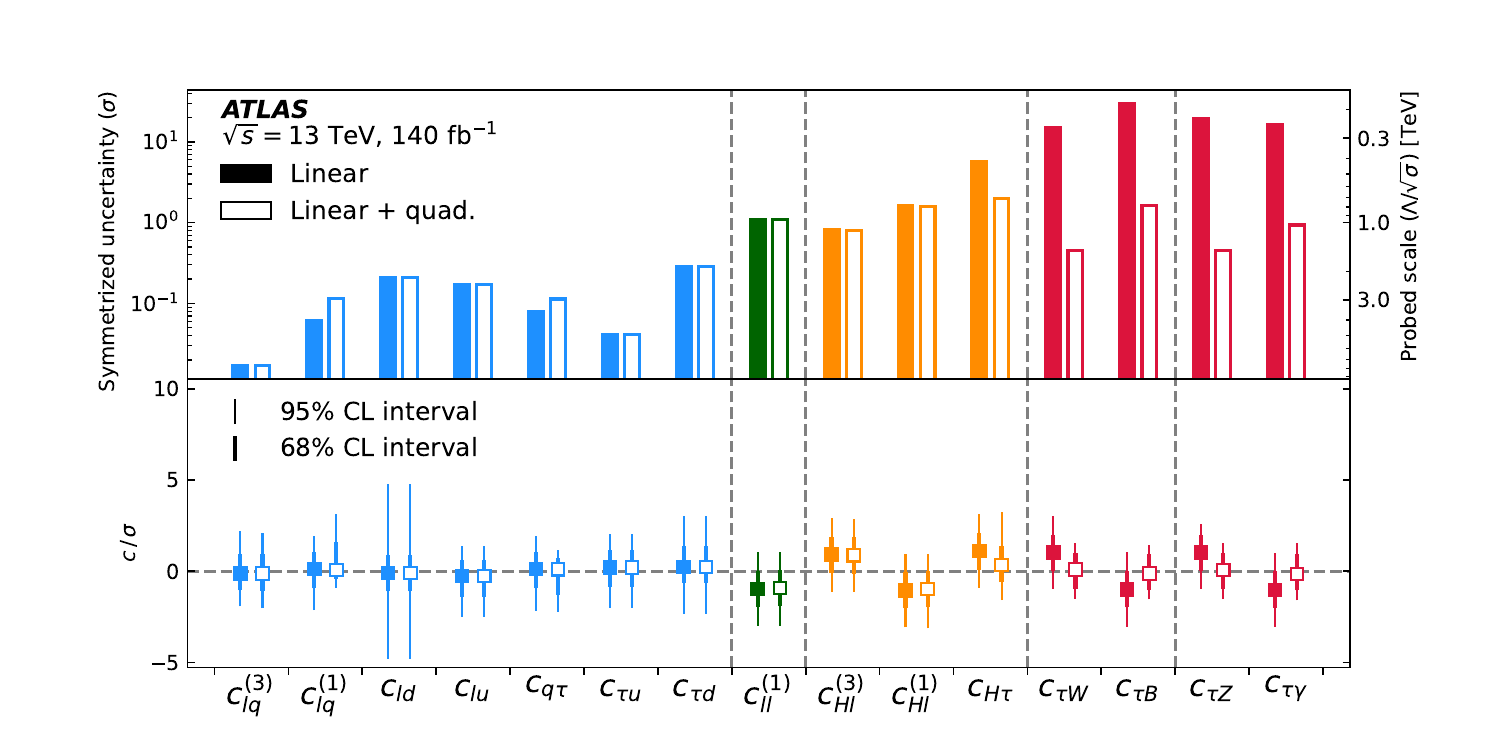}
\caption{ The results of likelihood fits for individual coupling coefficients using only the
first-order expansion of the coupling (`linear') or
the first-order and partial second-order contributions to the cross-section (`linear +
quad.').
The upper panel illustrates the symmetrized $1\sigma$ uncertainty in each
coupling coefficient at a scale of $\Lambda = 1$~TeV (left
$y$-axis) and the scale probed for a unit coefficient for the linear results (right
$y$-axis).
The lower panel shows the best-fit coefficient and the 68\% and 95\%
confidence intervals relative to the symmetrized $1\sigma$ uncertainty.
} \label{fig:eft-interp-summary}
\end{figure}

The results of the likelihood fits for individual SMEFT coefficients are given in
Table~\ref{tab:CoeffResults}; fits to the interference-only signal (`linear'),
as well as fits using both the interference and BSM terms (`linear + quad.'),
are shown.  The interference term dominates the sensitivity to all coefficients
except $c_{\tau B}$, $c_{\tau W}$, $c_{\tau Z}$, and $c_{\tau \gamma}$, for which
this term is suppressed and the BSM term dominates.
The best-fit value and 68\% confidence interval are given for each
coefficient at a scale of 1 TeV.  In the linear case the coefficient values scale
as $c(\Lambda) = c\Lambda^2/$TeV$^2$, so four-fermion interactions constrain
scales up to about 5~TeV for unit coefficients.
Constraints on effective $Z(/\gamma) \tau\tau$ interactions are weaker,
constraining the scale to more than 1~TeV for unit coefficients.
These uncertainties and constraints are illustrated in
Figure~\ref{fig:eft-interp-summary}.  Results of the $\chi^2$ fit to the particle-level
differential cross-section are consistent with the detector-level fits and in general
exhibit better constraining power than the detector-level fits for couplings that
do not enhance $b$-production, due to the finer binning in $\mtautauvis$.

\begin{table}[btp]
\centering
\caption{ Central values and 68\% confidence intervals from the individual coupling coefficient
fits when using the contribution linear in the coupling or using the linear and quadratic
contributions.  In the likelihood fits the scale is set to $\Lambda = 1$ TeV so the
values from the linear fit have the dependence $c(\Lambda) = c\Lambda^2/$TeV$^2$.
}
\label{tab:CoeffResults}
\begin{tabular}{c c c}
\toprule
Coefficient & fit result (linear) & fit result (linear + quad.) \\
\hline\interrowfill
\scalebox{0.8}{$c_{lq}^{(3)}$} & \scalebox{0.8}{$-0.002_{-0.016}^{+0.018}$} & \scalebox{0.8}{$-0.002_{-0.016}^{+0.018}$} \\ \interrowfill
\scalebox{0.8}{$c_{lq}^{(1)}$} & \scalebox{0.8}{$0.01_{-0.07}^{+0.06}$} & \scalebox{0.8}{$0.01_{-0.05}^{+0.18}$} \\ \interrowfill
\scalebox{0.8}{$c_{lu}$} & \scalebox{0.8}{$-0.05_{-0.20}^{+0.15}$} & \scalebox{0.8}{$-0.05_{-0.20}^{+0.15}$} \\ \interrowfill
\scalebox{0.8}{$c_{ld}$} & \scalebox{0.8}{$-0.02_{-0.21}^{+0.21}$} & \scalebox{0.8}{$-0.02_{-0.21}^{+0.21}$} \\ \interrowfill
\scalebox{0.8}{$c_{q\tau}$} & \scalebox{0.8}{$0.01_{-0.08}^{+0.07}$} & \scalebox{0.8}{$0.01_{-0.16}^{+0.07}$} \\ \interrowfill
\scalebox{0.8}{$c_{\tau u}$} & \scalebox{0.8}{$0.01_{-0.04}^{+0.04}$} & \scalebox{0.8}{$0.01_{-0.04}^{+0.04}$} \\ \interrowfill
\scalebox{0.8}{$c_{\tau d}$} & \scalebox{0.8}{$0.07_{-0.25}^{+0.32}$} & \scalebox{0.8}{$0.07_{-0.25}^{+0.32}$} \\ \interrowfill
\hline\interrowfill
\scalebox{0.8}{$c_{ll}^{(1)}$} & \scalebox{0.8}{$-1.1_{-1.1}^{+1.1}$} & \scalebox{0.8}{$-1.0_{-1.1}^{+1.1}$} \\ \interrowfill
\hline\interrowfill
\scalebox{0.8}{$c_{Hl}^{(3)}$} & \scalebox{0.8}{$0.7_{-0.8}^{+0.8}$} & \scalebox{0.8}{$0.7_{-0.8}^{+0.8}$} \\ \interrowfill
\scalebox{0.8}{$c_{Hl}^{(1)}$} & \scalebox{0.8}{$-1.7_{-1.6}^{+1.6}$} & \scalebox{0.8}{$-1.5_{-1.6}^{+1.6}$} \\ \interrowfill
\scalebox{0.8}{$c_{H\tau}$} & \scalebox{0.8}{$6_{-6}^{+6}$} & \scalebox{0.8}{$0.7_{-1.9}^{+2.1}$} \\ \interrowfill
\hline\interrowfill
\scalebox{0.8}{$c_{\tau W}$} & \scalebox{0.8}{$15_{-15}^{+15}$} & \scalebox{0.8}{$0.0_{-0.5}^{+0.4}$} \\ \interrowfill
\scalebox{0.8}{$c_{\tau B}$} & \scalebox{0.8}{$-30_{-30}^{+31}$} & \scalebox{0.8}{$-0.2_{-1.5}^{+1.8}$} \\ \interrowfill
\hline\interrowfill
\scalebox{0.8}{$c_{\tau Z}$} & \scalebox{0.8}{$20_{-20}^{+19}$} & \scalebox{0.8}{$0.0_{-0.5}^{+0.4}$} \\ \interrowfill
\scalebox{0.8}{$c_{\tau \gamma}$} & \scalebox{0.8}{$-17_{-16}^{+17}$} & \scalebox{0.8}{$-0.1_{-0.8}^{+1.0}$} \\ \interrowfill
\bottomrule
\end{tabular}
\end{table}

\begin{figure}[t!]
\centering
\includegraphics[width=0.95\textwidth]{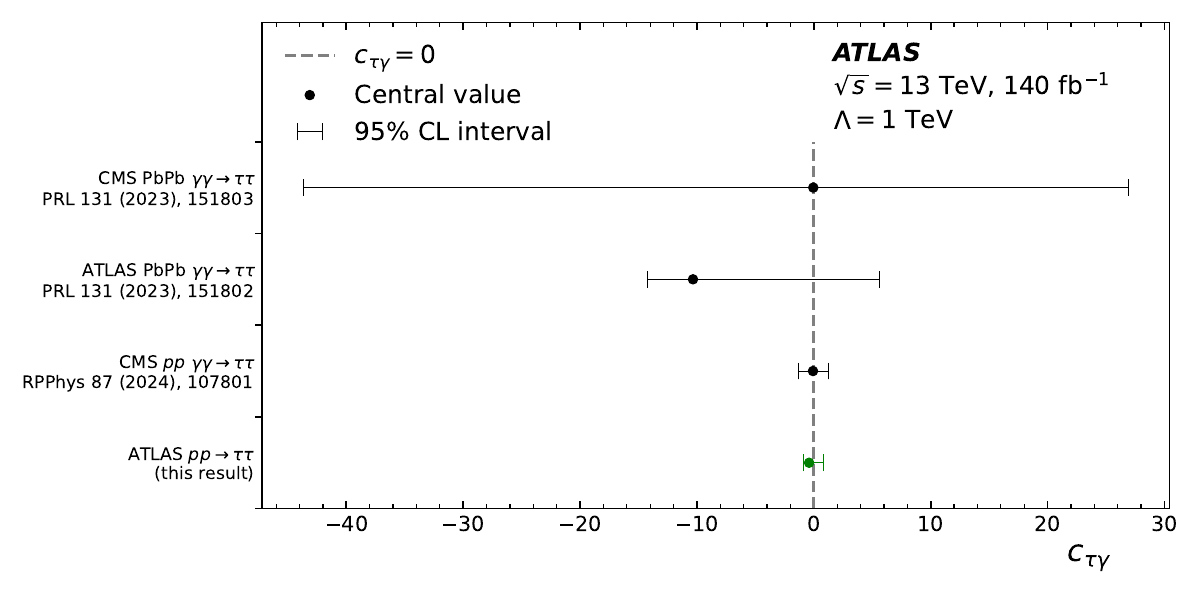}
\caption{
Comparison of central values and confidence intervals on $c_{\tau\gamma}$ from this measurement and
measurements probing the \taulepton anomalous magnetic moment~\cite{STDM-2019-19,CMS-SMP-23-005,CMS-HIN-21-009}.
} \label{fig:eft-interp-ctauy-comparison}
\end{figure}

As discussed in Section~\ref{sec:SMEFT}, $c_{\tau\gamma}$ affects the \taulepton anomalous magnetic
moment.  Previous LHC constraints on \atau~\cite{STDM-2019-19,CMS-SMP-23-005,CMS-HIN-21-009} can be
interpreted as constraints on $c_{\tau\gamma}$ using the EFT assumption that no new physics appears
at a low mass scale.  Figure~\ref{fig:eft-interp-ctauy-comparison} compares the results of these
measurements to that obtained from a $\chi^2$ fit to the particle-level differential cross-section
for $\tau\bar{\tau}$ production.  The confidence interval from this measurement is about a factor
of seven smaller than the ATLAS measurement in ultra-peripheral lead-lead collisions~\cite{STDM-2019-19},
and competitive with the CMS analysis of $\gamma\gamma \to \tau\bar{\tau}$ events in $pp$
collisions~\cite{CMS-SMP-23-005}.

The $c_{\tau\gamma}$ value of $-0.40_{-0.24}^{+0.63}$ from the $\chi^2$ fit can be translated into an \atau of
\atauwithuncert with a 95\% confidence interval of \ataurange using the relation in Section~\ref{sec:SMEFT}
and adding the SM value $\atau = 1.178 \times 10^{-3}$; however, since \atau is defined for on-shell photons
($q^2=0$) and this analysis relies on off-shell photon exchange, the interpretation should formally be
restricted to the EFT parameter $c_{\tau\gamma}$.  Nonetheless, one can apply the EFT assumption of no new
physics at a low scale and further infer \dtau from a $\chi^2$ fit for $c_{\tau\gamma}$ using the pure-SMEFT
contribution of the corresponding operator. This contribution has the same dependence on the real and imaginary
components of $c_{\tau\gamma}$ when it is complex, and there is negligible interference between the SM and
Im[$c_{\tau\gamma}$]-dependent matrix elements.  The resulting fit value and 95\% confidence interval are
\dtauwithuncert and \dtaurange, respectively.

\begin{figure}[t!]
\centering
\subfloat[]{
\includegraphics[width=0.49\textwidth]{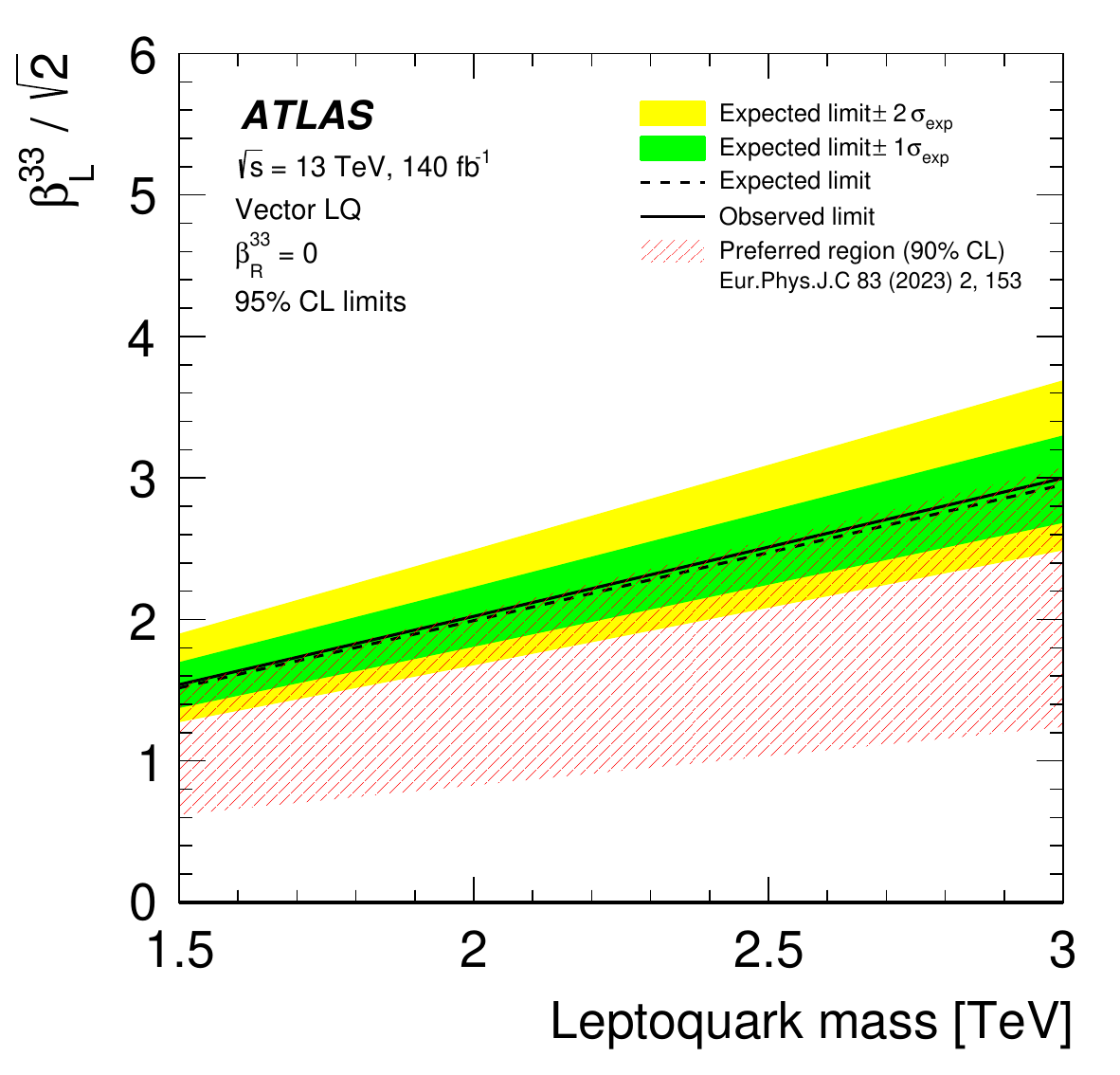}
}
\subfloat[]{
\includegraphics[width=0.49\textwidth]{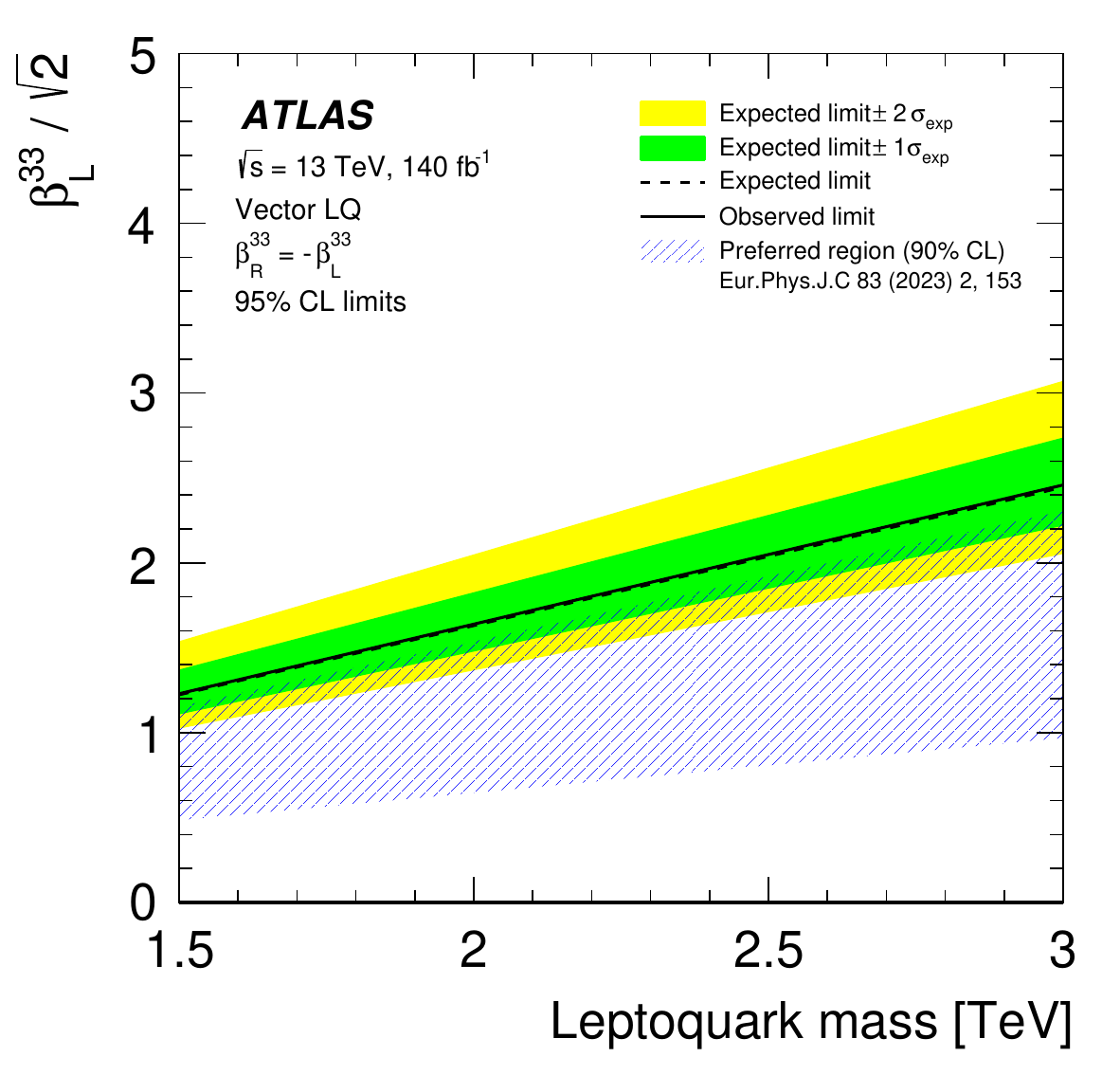}
}
\caption{ Coupling limits as a function of mass of a vector leptoquark coupling to a
$\tau$-lepton and a $b$ quark.
Shown are the observed (solid line) and expected limits (dashed line), along
with the expected 68\% (inner green band) and 95\% (outer yellow band) confidence
intervals, and the region preferred by measurements of $B$-hadron decays
(shaded).
Leptoquarks (a) coupling exclusively to left-handed fermions and (b)
coupling to left-handed and right-handed fermions with equal magnitude and
opposite sign are shown. }
\label{fig:VLQlimits}
\end{figure}

\begin{figure}[t!]
\centering
\subfloat[]{
\includegraphics[width=0.49\textwidth]{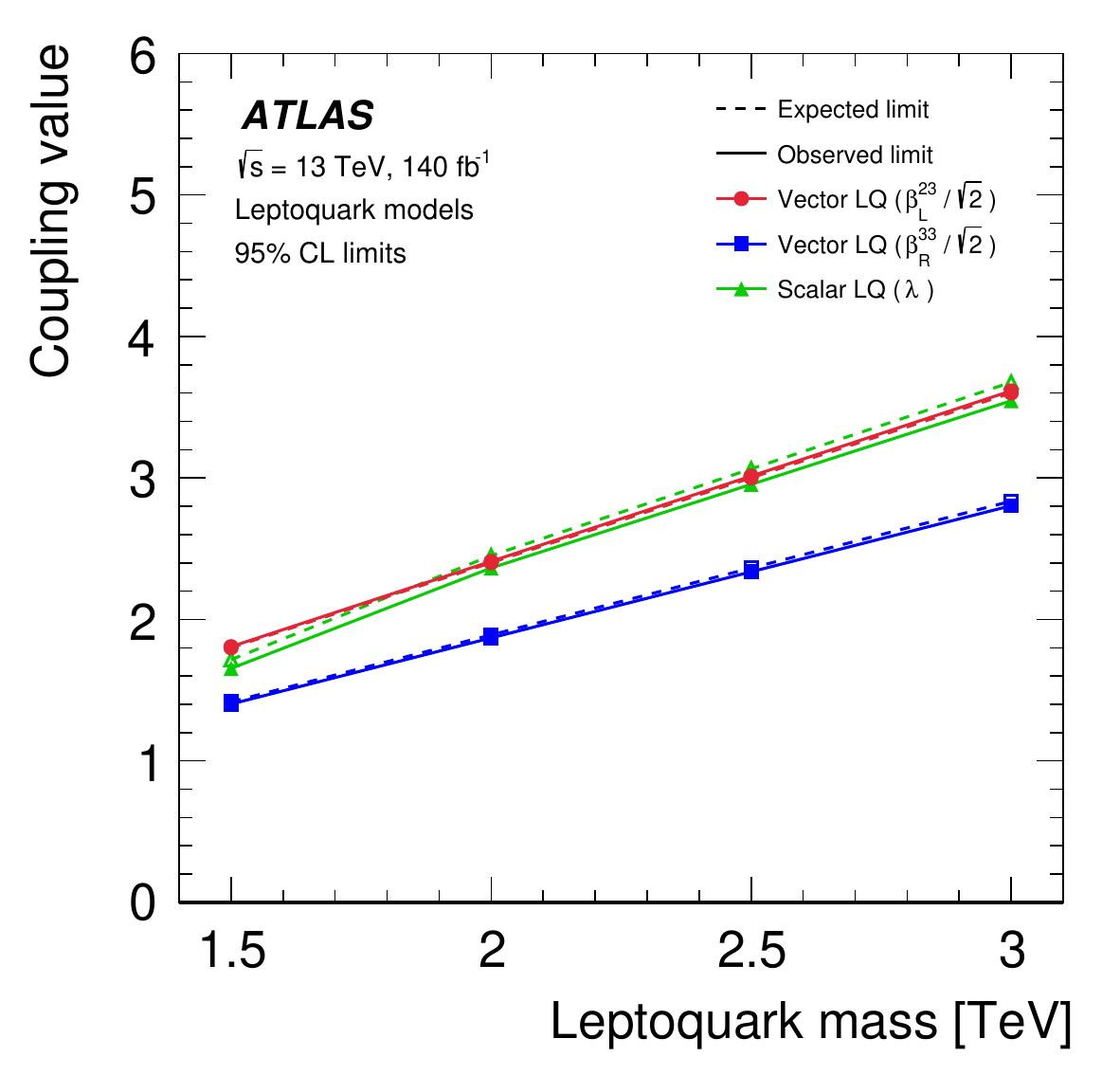}
}
\subfloat[]{
\includegraphics[width=0.49\textwidth]{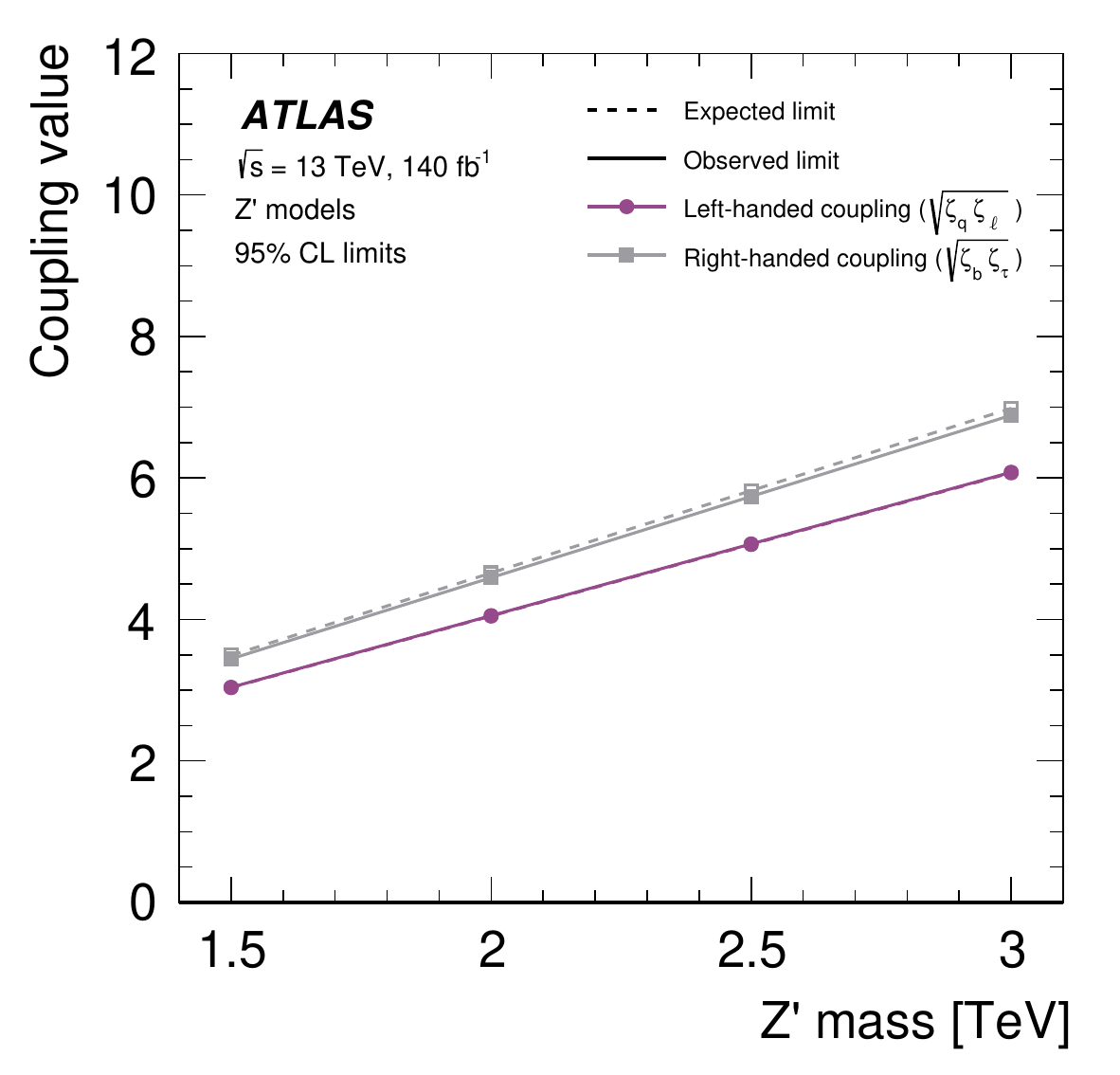}
}
\caption{(a) Coupling limits as a function of mass of a vector (red and blue) or scalar (green) leptoquark coupling to a
$\tau$-lepton and a $b$ quark.  Here the vector leptoquark couples to right-handed fermions (blue) or to left-handed
$\tau$-leptons and second generation quarks (red).
(b) The coupling limit as a function of mass for the non-resonant production of a $Z'$ boson coupling to left-handed
(dark line) or right-handed (light line) fermions. }
\label{fig:LQZprimelimits}
\end{figure}

No evidence of a signal is observed and limits are set on leptoquark and $Z'$-boson couplings as functions of their masses.
For the leptoquark coupling scenarios preferred by the $B$-hadron anomalies the limits exclude a small portion of
the preferred coupling-mass plane (Figure~\ref{fig:VLQlimits}).  Constraints are also set on the vector leptoquark couplings
$\beta_R^{33}$ and $\beta_L^{23}$, on the scalar leptoquark coupling $\lambda$, and on the $Z'$-boson coupling combinations
$\sqrt{\zeta_q \zeta_\ell}$ and $\sqrt{\zeta_b \zeta_\tau}$, all shown in Figure~\ref{fig:LQZprimelimits}.  The limits improve
on previous constraints from ATLAS~\cite{EXOT-2022-39}, and on those from CMS~\cite{CMS-EXO-19-016}.  The CMS search for
leptoquarks observed a local excess of 2.8 standard deviations that is most prominent for the non-resonant signal.  In events
without a $b$-jet CMS obtains a local significance between 3.4 and 3.7 standard deviations for the scalar and vector leptoquark
models, using a benchmark leptoquark mass of 2~TeV and coupling equal to 2.5.  This analysis excludes these benchmark parameter
values for scalar and vector leptoquark models.

\FloatBarrier

\section{Conclusion}
\label{sec:conclusions}

The first unfolded differential measurement of inclusive $\tau\bar{\tau}$ production has been performed
at the LHC, using 140 fb$^{-1}$ of proton--proton collision data recorded with the ATLAS detector.  The
measurement is consistent with the predictions of the Standard Model.  In addition, the data have been fit
directly for a variety of new non-resonant interactions.  Interactions involving quarks of the first two
generations are constrained using the SM effective field theory with a symmetry assumed between these
generations.  Interactions between a $\tau\bar{\tau}$ pair and a $Z$ boson or photon are also constrained.
Two models with enhanced couplings to third-generation fermions are constrained: those with a vector or
scalar leptoquark, and those with a $Z'$ boson.

Of particular interest are vector leptoquarks coupling to third-generation fermions with left-handed
couplings or a left-minus-right-handed combination.  Such couplings are preferred by $B$-hadron measurements
that show tension with SM predictions.  Sensitivity to these couplings is enhanced by separating the
data by $b$-jet multiplicity, since the most sensitive final state contains one additional $b$-jet.  The
combination of interference and leptoquark-only couplings are included for the first time in this final state.
The resulting constraints remove a small portion of the region preferred by the $B$-hadron measurements.

\section*{Acknowledgements}

%

%

%
%

%
%

We thank CERN for the very successful operation of the LHC and its injectors, as well as the support staff at
CERN and at our institutions worldwide without whom ATLAS could not be operated efficiently.

The crucial computing support from all WLCG partners is acknowledged gratefully, in particular from CERN, the ATLAS Tier-1 facilities at TRIUMF/SFU (Canada), NDGF (Denmark, Norway, Sweden), CC-IN2P3 (France), KIT/GridKA (Germany), INFN-CNAF (Italy), NL-T1 (Netherlands), PIC (Spain), RAL (UK) and BNL (USA), the Tier-2 facilities worldwide and large non-WLCG resource providers. Major contributors of computing resources are listed in Ref.~\cite{ATL-SOFT-PUB-2025-001}.

We gratefully acknowledge the support of ANPCyT, Argentina; YerPhI, Armenia; ARC, Australia; BMWFW and FWF, Austria; ANAS, Azerbaijan; CNPq and FAPESP, Brazil; NSERC, NRC and CFI, Canada; CERN; ANID, Chile; CAS, MOST and NSFC, China; Minciencias, Colombia; MEYS CR, Czech Republic; DNRF and DNSRC, Denmark; IN2P3-CNRS and CEA-DRF/IRFU, France; SRNSFG, Georgia; BMFTR, HGF and MPG, Germany; GSRI, Greece; RGC and Hong Kong SAR, China; ICHEP and Academy of Sciences and Humanities, Israel; INFN, Italy; MEXT and JSPS, Japan; CNRST, Morocco; NWO, Netherlands; RCN, Norway; MNiSW, Poland; FCT, Portugal; MNE/IFA, Romania; MSTDI, Serbia; MSSR, Slovakia; ARIS and MVZI, Slovenia; DSI/NRF, South Africa; MICIU/AEI, Spain; SRC and Wallenberg Foundation, Sweden; SERI, SNSF and Cantons of Bern and Geneva, Switzerland; NSTC, Taipei; TENMAK, T\"urkiye; STFC/UKRI, United Kingdom; DOE and NSF, United States of America.

Individual groups and members have received support from BCKDF, CANARIE, CRC and DRAC, Canada; CERN-CZ, FORTE and PRIMUS, Czech Republic; COST, ERC, ERDF, Horizon 2020, ICSC-NextGenerationEU and Marie Sk{\l}odowska-Curie Actions, European Union; Investissements d'Avenir Labex, Investissements d'Avenir Idex and ANR, France; DFG and AvH Foundation, Germany; Herakleitos, Thales and Aristeia programmes co-financed by EU-ESF and the Greek NSRF, Greece; BSF-NSF and MINERVA, Israel; NCN and NAWA, Poland; La Caixa Banking Foundation, CERCA Programme Generalitat de Catalunya and PROMETEO and GenT Programmes Generalitat Valenciana, Spain; G\"{o}ran Gustafssons Stiftelse, Sweden; The Royal Society and Leverhulme Trust, United Kingdom.

In addition, individual members wish to acknowledge support from Armenia: Yerevan Physics Institute (FAPERJ); CERN: European Organization for Nuclear Research (CERN DOCT); Chile: Agencia Nacional de Investigaci\'on y Desarrollo (FONDECYT 1230812, FONDECYT 1240864); China: Chinese Ministry of Science and Technology (MOST-2023YFA1605700, MOST-2023YFA1609300), National Natural Science Foundation of China (NSFC - 12175119, NSFC 12275265); Czech Republic: Czech Science Foundation (GACR - 24-11373S), Ministry of Education Youth and Sports (ERC-CZ-LL2327, FORTE CZ.02.01.01/00/22\_008/0004632), PRIMUS Research Programme (PRIMUS/21/SCI/017); EU: H2020 European Research Council (ERC - 101002463); European Union: European Research Council (BARD No. 101116429, ERC - 948254, ERC 101089007), European Regional Development Fund (SMASH COFUND 101081355, SLO ERDF), Horizon 2020 Framework Programme (MUCCA - CHIST-ERA-19-XAI-00), European Union, Future Artificial Intelligence Research (FAIR-NextGenerationEU PE00000013), Horizon 2020 (EuroHPC - EHPC-DEV-2024D11-051), Italian Center for High Performance Computing, Big Data and Quantum Computing (ICSC, NextGenerationEU); France: Agence Nationale de la Recherche (ANR-21-CE31-0022, ANR-22-EDIR-0002); Germany: Baden-Württemberg Stiftung (BW Stiftung-Postdoc Eliteprogramme), Deutsche Forschungsgemeinschaft (DFG - 469666862, DFG - CR 312/5-2); China: Research Grants Council (GRF); Italy: Istituto Nazionale di Fisica Nucleare (ICSC, NextGenerationEU), Ministero dell'Università e della Ricerca (NextGenEU I53D23000820006 M4C2.1.1); Japan: Japan Society for the Promotion of Science (JSPS KAKENHI JP22H01227, JSPS KAKENHI JP22H04944, JSPS KAKENHI JP22KK0227, JSPS KAKENHI JP23KK0245, JSPS KAKENHI JP24K23939); Norway: Research Council of Norway (RCN-314472); Poland: Ministry of Science and Higher Education (IDUB AGH, POB8, D4 no 9722), Polish National Science Centre (NCN 2021/42/E/ST2/00350, NCN OPUS 2023/51/B/ST2/02507, NCN OPUS nr 2022/47/B/ST2/03059, NCN UMO-2019/34/E/ST2/00393, UMO-2022/47/O/ST2/00148, UMO-2023/49/B/ST2/04085, UMO-2023/51/B/ST2/00920, UMO-2024/53/N/ST2/00869); Portugal: Foundation for Science and Technology (FCT); Spain: Ministry of Science and Innovation (MCIN \& NextGenEU PCI2022-135018-2, MICIN \& FEDER PID2021-125273NB, RYC2019-028510-I, RYC2020-030254-I, RYC2021-031273-I, RYC2022-038164-I); Sweden: Carl Trygger Foundation (Carl Trygger Foundation CTS 22:2312), Swedish Research Council (Swedish Research Council 2023-04654, VR 2021-03651, VR 2022-03845, VR 2022-04683, VR 2023-03403, VR 2024-05451), Knut and Alice Wallenberg Foundation (KAW 2018.0458, KAW 2022.0358, KAW 2023.0366); Switzerland: Swiss National Science Foundation (SNSF - PCEFP2\_194658); United Kingdom: Leverhulme Trust (Leverhulme Trust RPG-2020-004), Royal Society (NIF-R1-231091); United States of America: U.S. Department of Energy (ECA DE-AC02-76SF00515), Neubauer Family Foundation.

%
%


%
%
%
%
%
%

%
%

%
%
%
\printbibliography
\clearpage
 
\begin{flushleft}
\hypersetup{urlcolor=black}
{\Large The ATLAS Collaboration}

\bigskip

\AtlasOrcid[0000-0002-6665-4934]{G.~Aad}$^\textrm{\scriptsize 104}$,
\AtlasOrcid[0000-0001-7616-1554]{E.~Aakvaag}$^\textrm{\scriptsize 17}$,
\AtlasOrcid[0000-0002-5888-2734]{B.~Abbott}$^\textrm{\scriptsize 123}$,
\AtlasOrcid[0000-0002-0287-5869]{S.~Abdelhameed}$^\textrm{\scriptsize 119a}$,
\AtlasOrcid[0000-0002-1002-1652]{K.~Abeling}$^\textrm{\scriptsize 55}$,
\AtlasOrcid[0000-0001-5763-2760]{N.J.~Abicht}$^\textrm{\scriptsize 49}$,
\AtlasOrcid[0000-0002-8496-9294]{S.H.~Abidi}$^\textrm{\scriptsize 30}$,
\AtlasOrcid[0009-0003-6578-220X]{M.~Aboelela}$^\textrm{\scriptsize 45}$,
\AtlasOrcid[0000-0002-9987-2292]{A.~Aboulhorma}$^\textrm{\scriptsize 36e}$,
\AtlasOrcid[0000-0001-5329-6640]{H.~Abramowicz}$^\textrm{\scriptsize 157}$
% ...
\bigskip
\\

% ...
$^{17}$Department for Physics and Technology, University of Bergen, Bergen; Norway.\\
% ...
$^{30}$Physics Department, Brookhaven National Laboratory, Upton NY; United States of America.\\
% ...
$^{36}$$^{(a)}$Facult\'e des Sciences Ain Chock, Universit\'e Hassan II de Casablanca;$^{(b)}$Facult\'{e} des Sciences, Universit\'{e} Ibn-Tofail, K\'{e}nitra;$^{(c)}$Facult\'e des Sciences Semlalia, Universit\'e Cadi Ayyad, LPHEA-Marrakech;$^{(d)}$LPMR, Facult\'e des Sciences, Universit\'e Mohamed Premier, Oujda;$^{(e)}$Facult\'e des sciences, Universit\'e Mohammed V, Rabat;$^{(f)}$Institute of Applied Physics, Mohammed VI Polytechnic University, Ben Guerir; Morocco.\\
% ...
$^{45}$Physics Department, Southern Methodist University, Dallas TX; United States of America.\\
% ...
$^{49}$Fakult\"{a}t Physik , Technische Universit{\"a}t Dortmund, Dortmund; Germany.\\
% ...
$^{55}$II. Physikalisches Institut, Georg-August-Universit\"{a}t G\"ottingen, G\"ottingen; Germany.\\
% ...
$^{104}$CPPM, Aix-Marseille Universit\'e, CNRS/IN2P3, Marseille; France.\\
% ...
$^{119}$$^{(a)}$New York University Abu Dhabi, Abu Dhabi;$^{(b)}$United Arab Emirates University, Al Ain; United Arab Emirates.\\
% ...
$^{123}$Homer L. Dodge Department of Physics and Astronomy, University of Oklahoma, Norman OK; United States of America.\\
% ...
$^{157}$Raymond and Beverly Sackler School of Physics and Astronomy, Tel Aviv University, Tel Aviv; Israel.\\
% ...
$^{a}$ Also at Affiliated with an institute formerly covered by a cooperation agreement with CERN.\\
$^{b}$ Also at An-Najah National University, Nablus; Palestine.\\
$^{c}$ Also at Borough of Manhattan Community College, City University of New York, New York NY; United States of America.\\
$^{d}$ Also at Center for Interdisciplinary Research and Innovation (CIRI-AUTH), Thessaloniki; Greece.\\
$^{e}$ Also at Centre of Physics of the Universities of Minho and Porto (CF-UM-UP); Portugal.\\
$^{f}$ Also at CERN, Geneva; Switzerland.\\
% ...

\end{flushleft}


%

%
%

%

%
%
%
%

\end{document}